\documentclass[iop]{emulateapj}
\usepackage{color}
\usepackage{natbib}
\usepackage{ulem}
\usepackage{amssymb,amsmath}

\begin{document}
\shorttitle{Habitable Zone of S-Type Binary Star Systems}
\shortauthors{Kaltenegger \& Haghighipour}

\title{Calculating the Habitable Zone of Binary Star Systems I: S-Type Binaries}
\author{Lisa Kaltenegger\altaffilmark{1,2} and Nader Haghighipour\altaffilmark{3,4}}

\altaffiltext{1}{MPIA, Koenigstuhl 17, Heidelberg, D 69117, Grmany}
\altaffiltext{2}{CfA, MS-20, 60 Garden Street, Cambridge, MA 02138, USA}
\altaffiltext{3}{Institute for Astronomy and NASA Astrobiology Institute,
University of Hawaii-Manoa, Honolulu, HI 96822}
\altaffiltext{4}{Institute for Astronomy and Astrophysics, University of Tuebingen, 
72076 Tuebingen, Germany}
\email{kaltenegger@mpia.de}

\begin{abstract}

We have developed a comprehensive methodology for calculating the boundaries of the 
habitable zone (HZ) of planet-hosting S-type binary star systems. Our approach is general 
and takes into account the contribution of both stars to the location and extent of the 
Binary HZ with different stellar spectral types. We have studied how the binary eccentricity 
and stellar energy distribution affect the extent of the habitable zone. Results indicate 
that in binaries where the combination of mass-ratio and orbital eccentricity allows planet 
formation around a star of the system to proceed successfully, the effect of a less luminous 
secondary on the location of the primary's habitable zone is generally negligible. However, 
when the secondary is more luminous, it can influence the extent of
the HZ. We present the details of the derivations of our methodology
and discuss its application to the Binary HZ around the primary and secondary main sequence
stars of an FF, MM, and FM binary, as well as two known planet-hosting binaries $\alpha$ Cen AB and HD 196886. 

\end{abstract}

\keywords{Astrobiology: Habitable zone -- Stars: binaries -- Stars: Planetary systems -- atmospheric effects}

\section{Introduction}

The discovery of circumstellar planets in close binary systems (i.e. binaries with stellar 
separations smaller than 50 AU) in the past two decades has lent strong support to the fact 
that planet formation around a star of a binary is robust, and these systems (known as S-type 
binaries, figure 1) can host a variety of planets including small, terrestrial-class objects 
(see Haghighipour 2010a\&b for a review). At present, there are six close S-type systems 
that host planets; GL 86 (Queloz et al. 2000; Eggenberger et al. 2001), $\gamma$ Cephei 
(Hatzes et al. 2003; Endl et al. 2011), HD 41004 (Zucker et al. 2004), HD 196885 
(Correia et al. 2008, Chauvin et al. 2011), 
HD 176051 (Muterspaugh et al. 2010), and $\alpha$ Centauri (Dumusque et al. 2012, we note that
the existence of the planet around the secondary star of this system has recently been challenged by
Hatzes 2013). Although Earth-like planets are yet to be discovered 
in the habitable zone (HZ) of binary star systems, studies of the formation and long-term 
stability of these objects have shown that depending on the binary semimajor axis, eccentricity, 
and mass-ratio, Earth-sized planets can form around a star of a binary in a close S-type system 
and can have stable orbits in the star's HZ (Haghighipour 2006; Quintana et al. 2007; Haghighipour \& Raymond 2007; 
Guedes et al. 2008; Th\'ebault, Marzari \& Scholl 2008, 2009; Eggl, Haghighipour \& Pilat-Lohinger 2013, 
Eggl, et al. 2013). 

In all these simulations, it has been generally assumed that the HZ of the binary is equivalent 
to the single-star HZ of its planet-hosting star. Although in binaries with separations smaller 
than 50 AU, the secondary star plays an important role in the formation, long-term stability, 
and water content of a planet in the HZ of the primary, 
the effect of the secondary on the range and extent of the HZ 
in these systems was ignored [in S-type systems, the motion of the 
binary around its center of mass is neglected, and the secondary star is considered to orbit a stationary
primary (bottom panel of figure 1)]. However, the fact that this star can affect planet formation around 
the primary, and can also perturb the orbit of a planet in the primary's HZ in binaries with moderate 
eccentricities, implies that the secondary may play a non-negligible role in the habitability of 
the system as well. In this paper, we present a methodology for calculating the HZ of S-type 
binaries by taking the effect of both stars into account, as well as calculate the stability of a 
planet in the HZ.

Unlike around single stars where the HZ is a spherical shell with a distance determined by the 
host star alone, in binary star systems, the radiation from the stellar companion can influence 
the extent and location of  the HZ of the system. Especially for planet-hosting binaries with small
stellar separations and/or in binaries where the planet orbits the less luminous star, the amount 
of the flux received by the planet from the secondary star may become non-negligible. 

In addition, effects such as the gravitational perturbation of the secondary star 
(see e.g., Georgakarakos 2002, Eggl et al. 2012) 
can influence a planet's orbit in the Binary HZ and lead to temperature fluctuations if the planetary 
atmosphere cannot buffer the change in the combined insolation. Since in an S-type system, the secondary 
orbits more slowly than the planet, during one period of the binary, the planet may experience the effects 
of the secondary several times. The latter, when combined with the atmospheric response of the planet, 
defines the Binary HZ of the system. In this paper we concentrate on the extent of the Binary HZ, and not the 
dynamical effect of a binary on the orbit of individual planets which depends on specific system 
parameters.

Despite the fact that as a result of the orbital architecture and dynamics of the binary, at times 
the total radiation received by the planet exceeds the radiation that it receives from its parent 
star alone by a non-negligible amount, the boundaries of the actual HZ of the binary cannot be obtained 
by a simple extrapolation of the boundaries of the HZ of its planet-hosting star. Similar to the HZ around 
single stars, converting from insolation to equilibrium temperature of the planet depends strongly 
on the planet's atmospheric composition, cloud fraction, and star's spectral type. A planet's 
atmosphere responds differently to stars with different spectral distribution of incident 
energy. Different stellar types will therefore contribute differently to the total amount of energy 
absorbed by the planetary atmosphere (see e.g., Kasting et al. 1993). A complete and realistic 
calculation of the HZ has to take into account the spectral energy distribution (SED) of the binary 
stars as well as the planet's atmospheric response. In this paper, we address these issues and present 
a coherent and self-consistent model for determining the boundaries of the HZ of S-type binary systems. 

We describe our model and present the calculations of the HZ in Section 2. In section 3.1, 
we calculate the maximum flux of the secondary on the single-star HZ
of the primary in three general binary systems with F-F, M-M, and F-M stars as examples. In sections 3.2 and 3.3, we then
demonstrate how to correctly estimate the Binary HZ taking into account the contribution of both 
components of the binary as well as the stability constraints. We then apply our methodology to
two known exoplanetary systems, $\alpha$ Centauri and HD 196885.
Among the currently known moderately close planet hosting S-type binaries,
these two systems are the only ones with main sequence stars and known stellar characteristics 
(the primary star of the GL 86 system is a white dwarf, that of $\gamma$ Cephei system is a K IV
sub-giant, the HD41004 system is a hierarchical quadruple system, and it is unclear which 
star in the binary HD 176051 hosts its planet).  
We do not take the known planets in these system into account and instead consider them to host a 
fictitious Earth-like planet with a CO$_2$/H$_2$O/N$_2$ atmosphere (following Kasting et al. 1993, 
Selsis et al. 2007, Kaltenegger \& Sasselov 2011, and Kopparapu et al. 2013a). 
We calculate the HZ of the binary for the cases where the Earth-like planet orbits the primary or
the secondary star, and study the effect of the binary eccentricity on the width and location of 
the Binary HZ. In section 4, we discuss the effect of planet eccentricity, and in section 5, 
we conclude this study by summarizing the results and discussing their implications.

\section{Description of the Model and Calculation of the Binary Habitable Zone}

Habitability and the location of the HZ depend on the stellar flux at the planet's location as well 
as the planet's atmospheric composition. The latter determines the albedo and the greenhouse effect 
in the planet's atmosphere and as such plays a strong role in determining the boundaries of the HZ. 
Examples of atmospheres with different chemical compositions include the original CO$_2$/H$_2$O/N$_2$ 
model (Kasting et al 1993; Selsis et al. 2007; Kopparapu et al. 2013a) with a water reservoir like 
Earth's, and model atmospheres with high H$_2$/He concentrations (Pierrehumbert \& Gaidos 2011) or 
with limited water supply (Abe et al. 2011).
 
In this paper, we use the recent update to the Sun's HZ model (Kopparapu et al. 2013a\&b). According 
to this model, the HZ is an annulus around a star where a rocky planet with a CO$_2$/H$_2$O/N$_2$ atmosphere 
and sufficiently large water content (such as on Earth) can host liquid water permanently on its 
solid surface (which allows remote detectability of atmospheric biosignatures). This definition of the 
HZ assumes the abundance of CO$_2$ and H$_2$O in the atmosphere is regulated by a geophysical cycle similar 
to Earth's carbonate silicate cycle. The inner and outer boundaries of the HZ in this model are associated 
with a H$_2$O-- and CO$_2$--dominated atmosphere, respectively. Between those limits on a geologically active 
planet, climate stability is provided by a feedback mechanism in which atmospheric CO$_2$ concentration varies 
inversely with planetary surface temperature. 

The locations of the inner and outer boundaries of a single star's as well as a binary's HZ depend also 
on the cloud fraction in the planet's atmosphere. That is because the overall planetary albedo $A$ is a 
function of the chemical composition of the clear atmosphere as well as the additional cooling or warming 
of the atmosphere due to clouds ($A=A_{\rm clear}+A_{\rm cloud}$). 
In this paper, we use the region between runaway and maximum
greenhouse limits from the recent HZ model as the {\it narrow} HZ
(Kopparapu et al 2013a\&b). This model does not include cloud feedback. 
Therefore, we use the {\it empirical} HZ as a second limit that is derived 
using the fluxes received by Mars and Venus at 3.5 and 1.0 Gyr, respectively (the region between Recent Mars
and Early Venus). At these times, the two planets do not show indications for liquid water on their surfaces 
(see Kasting et al. 1993).
In this definition, the locations of the HZs are determined based on the flux received by the planet 
(see e.g., Kasting et al. 1993; Selsis et al. 2007; Kaltenegger \& Sasselov 2011; and Kopparapu et al. 2013a).

\subsection{Effect of Star's Spectral Energy Distribution (SED)}

The locations of the boundaries of the HZ depend on the flux of the star at the orbit of the planet. 
In a binary star system where the planet is subject to radiation from two stars, the flux of the 
secondary star has to be added to that of the primary (planet-hosting star) and the total flux 
can then be used to calculate the boundaries of the Binary HZ. 
However, because the response of a planet's atmosphere to the radiation from 
a star depends strongly on the star's SED, a simple summation of fluxes is not applicable. The absorbed 
fraction of the absolute incident flux of each star at the top of the planet's atmosphere will differ 
for different SEDs. The planet's Bond albedo increases with the star's effective temperature 
$(T_{\rm Star})$, because for stars with higher values of $T_{\rm Star}$, more stellar photons are deposited 
at the top of planet's atmosphere in the short wavelengths where Rayleigh scattering in a planet's 
atmosphere is very efficient (see the definition of the spectral weight factor in section 2.2). 
That increases the amount of reflected stellar light for hotter stars. 
As a result, the absolute incident stellar flux at the top of the planetary atmosphere that leads 
to a similar absorbed stellar flux and surface temperature for similar planetary atmospheres 
is larger for hotter stars (see section 2.2 and figure 2). 
Therefor in order to add the absorbed flux of two different stars and derive 
the limits of the Binary HZ, one has to weight the flux of each star according to the 
star's SED. The relevant flux received by a planet in this case is the sum of the  
spectrally weighted stellar flux, separately received from each star of the binary, as given by equation (1),

\begin{equation}
\begin{split}
{F_{\rm Pl}}(f,{T_{\Pr}},{T_{\rm Sec}})\,=\,
{W_{\rm Pr}}(f,{T_{\rm Pr}})\, {{{L_{\rm Pr}}({T_{\rm Pr}})}\over {r_{\rm Pl-Pr}^2}} \\
 +\, {W_{\rm Sec}}(f,{T_{\rm Sec}})\, {{{L_{\rm Sec}}({T_{\rm Sec}})}\over {r_{\rm Pl-Sec}^2}} 
\end{split}
\end{equation}

\noindent
In this equation, $F_{\rm Pl}$ is the total flux received by the planet, $L_i$ and $T_i$ ($i$=Pr, Sec) 
represent the luminosity and effective temperature of the primary and secondary stars, $f$ is the cloud 
fraction of the planet's atmosphere, and ${W_i}(f,{T_i})$ is the binary stars' spectral weight factor. 
The quantities $r_{\rm Pl-Pr}$ and $r_{\rm Pl-Sec}$ in equation (1) represent the distances between the 
planet and the primary and secondary stars, respectively. In using equation (1), we normalize the 
weighting factor to the flux of the Sun.

From equation (1), the boundaries of the HZ of the binary can be defined as distances where the total 
flux received by the planet is equal to the flux that Earth receives from the Sun at the inner and 
outer edges of its HZ. Since in an S-type system, the planet revolves around one star of the binary, 
we determine the inner and outer edges of the HZ with respect to the planet-hosting star. 
As mentioned before, it is customary to consider the primary of the system to be stationary, 
and calculate the orbital elements with respect to the stationary primary star (see bottom panel of figure 1). 
In the rest of this paper, we will follow this convention and consider the planet-hosting star to be
the primary star as well. In that case, the range of the HZ of the binary can be obtained from

\begin{equation}
{W_{\rm Pr}}(f,{T_{\rm Pr}})\, {{{L_{\rm Pr}}({T_{\rm Pr}})}\over {l_{\rm x-Bin}^2}}+
{W_{\rm Sec}}(f,{T_{\rm Sec}})\, {{{L_{\rm Sec}}({T_{\rm Sec}})}\over {r_{\rm Pl-Sec}^2}}=
{{L_{\rm Sun}}\over {l_{\rm x-Sun}^2}}
\end{equation}

\noindent
In equation (2), the quantity 
${l_{\rm x}}$ represents the inner and outer edges of the HZ with x=(in,out). As mentioned earlier, the 
values of $l_{\rm in}$ and $l_{\rm out}$ are model-dependent and change for different values of cloud 
fraction, $f$, and atmosphere composition.

\subsection{Calculation of Spectral Weight Factors}

To calculate the spectral weight factor $W(f,T)$ for each star of the binary depending on their SEDs, 
we calculate the stellar flux at the 
top of the atmosphere of an Earth-like planet at the limits of the HZ, in terms of the stellar effective 
temperature. To determine the locations of the inner and outer boundaries of the HZ of a main sequence star 
with an effective temperature of $2600 K < {T_{\rm Star}} < 7200$ K, we use equation (3) (see Kopparapu et al 2013a)

\begin{equation}
{l_{\rm x-Star}}\,=\,{l_{\rm x-Sun}}\, 
\Biggl[{{L/{L_{\rm Sun}}}\over{1+{\alpha_{\rm x}} ({T_i})\,{l_{\rm x-Sun}^2}}} \Biggr]^{1/2}
\end{equation}

\noindent
In this equation, ${l_{\rm x}}=({l_{\rm in}},{l_{\rm out}})$ is in AU, ${T_i}{\rm (K)}={T_{\rm Star}}{\rm (K)}-5780$,
and 

\begin{equation}
{\alpha_{\rm x}}({T_i})\,=\,{a_{\rm x}}{T_i}\,+\,{b_{\rm x}}{T_i^2}\,+\,{c_{\rm x}}{T_i^3}\,+\,{d_{\rm x}}{T_i^4}
\end{equation}

\noindent
where the values of coefficients ${a_{\rm x}}, {b_{\rm x}}, {c_{\rm x}}$, ${d_{\rm x}}$, and $l_{\rm x-Sun}$ are given 
in Table 1 (see Kopparapu et al. 2013b). From equation (3), the flux received by the planet from its host star at the 
limits of the Habitable Zone can be calculated using equation (5). The results are given in Table 1;

\begin{equation}
{F_{\rm x-Star}}\big(f, {T_{\rm Star}}\big)\,=\,{F_{\rm x-Sun}}(f,{T_{\rm Star}})\,
\bigg[1\,+\,{\alpha_{\rm x}}({T_i})\,{l_{\rm x-Sun}^2}\bigg]
\end{equation}

\noindent
From equation (5), the spectral weight factor $W(f,T)$ can be written as

\begin{equation}
{W_i}(f,{T_i})\,=\,\biggl[1\,+\,{\alpha_{\rm x}}({T_i})\, {l_{\rm x-Sun}^2}\,\biggr]^{-1}
\end{equation}

\noindent
Table 2 and figure 3 show $W(f,T)$ as a  function of the effective temperature of a 
main sequence planet-hosting star for the narrow (top panel) and empirical (bottom panel) boundaries of the HZ. 
As expected, hotter stars have weighting factors smaller than 1 whereas the weighting factors of cooler 
stars are larger 1.

\subsection{Effect of Binary Eccentricity}

To use equation (2) to calculate the boundaries of the HZ, we assume here that the orbit of the 
(fictitious) Earth-like planet around its host star is circular. In a close binary system, the gravitational 
effect of the secondary may deviate the motion of the planet from a circle and cause its orbit to become eccentric. 
In a binary with a given semimajor axis and mass-ratio, the eccentricity has to stay within a 
small to moderate level to avoid strong interactions between the secondary star and the planet, and to allow the 
planet to maintain a long-term stable orbit (with a low eccentricity) in the primary's HZ. The binary eccentricity 
itself is constrained by the fact that in highly eccentric systems, periodic close approaches of the two stars
truncate their circumstellar disks depleting them from planet-forming material (Artymowicz \& Lubow 1994) and
restricting the delivery of water-carrying objects to an accreting
terrestrial planets in the Binary HZ (Haghighipour \& Raymond 2007).

This all indicates that in order for the binary to be able to form a terrestrial planet in its HZ, 
its eccentricity cannot have large values. In a binary with a small eccentricity, the deviation of the 
planet's orbit from circular is also small and appears in the form of secular changes with long periods 
(see e.g. Eggl et al. 2012). Therefore, to use equation (2), one can approximate the orbit of the planet 
by a circle without the loss of generality.
      
The habitability of a planet in a binary system also requires long-term stability in the HZ. 
For a given semimajor axis ${a_{\rm Bin}}$, eccentricity ${e_{\rm Bin}}$, and mass-ratio $\mu$ of the 
binary, there is an upper limit for the semimajor axis of the planet beyond which the perturbing effect 
of the secondary star will make the orbit of the planet unstable. This maximum or {\it critical} semimajor axis 
$({a_{\rm Max}})$ is given by (Rabl \& Dvorak 1988, Holman \& Wiegert 1999)

\begin{equation}
\begin{split}
{a_{\rm Max}}\,=\,{a_{\rm Bin}}\Big(0.464 - 0.38\, \mu - 0.631\, {e_{\rm Bin}} + 
0.586 \, \mu \, {e_{\rm Bin}} \\
+ 0.15 \, {e_{\rm Bin}^2} - 0.198\, \mu \,{e_{\rm Bin}^2}\Big)
\end{split}
\end{equation} 

\noindent
In equation (7), $\mu = {m_2}/({{m_1}+{m_2}})$ where ${m_1}$ and $m_2$ are the masses of the primary 
(planet-hosting) and secondary stars, respectively. 
One can use equation (7) to determine the maximum binary eccentricity that would allow the planet to 
have a stable orbit in the HZ (${l_{\rm out}} \leq {a_{\rm Max}}$). For any smaller value of the binary eccentricity, 
the entire HZ will be stable. 
We will present a detailed discussion of this topic in sections 3.2 and 3.3 where we calculate 
the boundaries of the HZ of  the $\alpha$ Centauri and HD 196885 systems, respectively.

\section{The Habitable Zone of Main Sequence S-Type Binaries}

Without knowing the exact orbital configuration of the planet, one can only estimate 
the boundaries of the Binary HZ by calculating the maximum and minimum additional 
flux from the secondary star at its closest and furthest distances from a fictitious Earth-like planet, 
as a first order approximation. This brackets the limits of the Binary HZ. Note that using the maximum flux of the 
secondary onto the planet for calculating the new Binary HZ overestimates the shift of the HZ from the single 
star's HZ to the Binary HZ due to the secondary because the planet's atmosphere can buffer an increase in radiation 
temporarily. This shift is underestimated when one uses the minimum flux received from the secondary star onto the planet.
To improve on this estimation, one needs to know the orbital positions of the planet as well as the stars in the binary. 
That way one can determine the exact flux over time reaching the planet as well as the number of planetary orbits over 
which the secondary's flux can be averaged. This depends on the system's geometry (both stellar and planetary parameters) 
and needs to be calculated for each planet hosting S-type system, individually. We assume here that the orbit of the planet around its host star is circular.

\subsection{Influence of the secondary on the single-star HZ of the primary}
To explore the maximum effect of the binary semimajor axis and eccentricity on the contribution 
of one star to the extent of the HZ around the other component, we consider three extreme cases: an M2-M2 (Figure 4, top), 
an F0-F0 (Figure 4, bottom), and an F8-M1 (Figure 5) binary. 
We consider the M2 and F0 stars to have effective temperatures of 3520 (K) and 7300 (K), respectively, and their 
luminosities to be 0.035 and 6.56 of that of the Sun for our general examples here. 

We first study the effect of a secondary on the single star HZ around the primary star. Figure 4 and Figure 5 show the 
maximum contribution of the secondary to the stellar flux at the limits of the single star's HZ of the primary, when the 
secondary star and a fictitious rocky planet are at their closest separation, and for different values of the binary 
eccentricity and semimajor axis. The color coding in these figures corresponds to the contribution of the flux received 
from the secondary relative to that of the primary star (note the multiplication factor on the lower right corner of each panel). 
Figure 4 shows secondary's flux at the inner (left) and outer (right) edge of the primary's single star empirical HZ, for a M2-M2 binary (top) and for a F0-F0 binary (bottom). The contribution of the secondary can be several times larger than the primary's at closest approach. The contribution of the secondary is larger at the outer edge of the primary's HZ than at the inner edge and increases with decreasing periastron distance of the binary.  Note that the actual flux contribution of the secondary that determines the Binary HZ limits is the secondary's flux averaged over the planet's full orbit  and as such it will be smaller than the secondary's flux at closest approach.

To explore the effect of the secondary for a binary with a hot and a cool star, we use an F-M system (see Figure 5). The top panels of Figure 5 show the maximum flux of the M1V secondary star at the outer edges of the F8V primary's single star narrow (left) and empirical HZs (right) when the secondary  is at its closest distance (i.e., during its periastron passage). The bottom panels show the maximum flux of the F8V at the outer edges of the single star narrow and empirical HZ of the M1V star. As shown by these panels, the flux of the brighter star has a stronger contribution to the total flux at the outer edge of the primary's HZ  (up to several times the primary's flux). Figure 4 and Figure 5 do not consider the orbital (in)stability of the fictitious Earth-like planet. When in an individual case, the stability criterion, as shown in equation (7), is imposed, the closest distance between the secondary and planet, which ensures the orbital stability of the planet as well, will be larger than the closest distance shown in these figures, and as a result, the contribution of the secondary star to the total flux received by the planet is smaller.

As a second step, to demonstrate the effect of the secondary on the boundaries of the HZ, we calculate the
Binary HZ of the systems mentioned above, considering the minimum value of the binary semimajor axis for which
the outer edge of the primary's empirical HZ will be on the stability limit. Figures 6 shows the results
for the case of the three binaries described above, assuming circular orbits, (top) an M2-M2 (left), an F0-F0 (right) and (bottom) an F8-M1 S-type Binary, with (left) the F8 star and (right) the M1 star being the planet hosting star. The top panel of Fig. 6 shows that the secondary does not have a noticeable effect on the extent of the HZ. The Binary HZ around each star is equivalent to its single-star HZ for the M2-M2 and F0-F0 S-type systems when stability up to the outer region of the empirical HZ is maintained. Also as expected, the effect of the M1 star on the extension of the single-star HZ around the F8 star is negligible. However, at its closest distances, the F8 star can extent the limit of the single-star HZ around the M1 star so far out that at the binary periastron, the two HZs merge. 

To further explore the effect of binary eccentricity in a system with a hot and cool star, we carried out similar
calculations as those in Figure 6, for the F8-M1 binary, assuming the binary eccentricity to be 0.3. Figures
7 and 8 show the results for four different relative positions of the two stars to show the changing influence of the secondary over the binary's orbit. The primary and planet hosting star is chosen to be the M star (Figure 7) and the F star (Figure 8). Figure 7 shows that a luminous secondary can have considerable effects on the shape and location of the HZ. However,
a cool and less luminous secondary will not change the limits of the HZ (Figure 8). The figures also demonstrate why the exact boundaries of the Binary HZ can only be calculated when the relative position of the planet to the stars is known - and the flux received by the planet over one full planetary orbit can be calculated.

\subsection{Binary HZ - Example: $\alpha$ Centauri}
The $\alpha$ Centauri system consists of the close binary $\alpha$ Cen AB, and a farther M dwarf companion 
known as Proxima Centauri at approximately 15000 AU. The semimajor axis of the binary is $\sim 23.5$ AU 
and its eccentricity is $\sim 0.518$. The component A of this system is a G2V star with a mass of $1.1\, {M_{\rm Sun}}$, 
luminosity of $1.519\,{L_{\rm Sun}}$, and an effective temperature of 5790 K. Its component B has a spectral 
type of K1V, and its mass, luminosity and effective temperature are equal to 
$0.934\, {M_{\rm Sun}}$, $0.5\,{L_{\rm Sun}}$, and 5214 K, respectively.

The recent announcement of a probable super-Earth planet with a mass larger than 1.13 Earth-masses around 
$\alpha$ Cen B indicates that unlike the region around $\alpha$ Cen A where terrestrial planet formation 
encounters complications, planet formation is efficient around this star 
(Guedes et al. 2008, Th\'ebault, Marzari \& Scholl 2009)
and it could also host a terrestrial planet in its HZ. Here we assume that planet formation around both 
stars of this binary can proceed successfully and they both can host Earth-like planets. We calculate the 
spectral weight factors of both $\alpha$ Cen A and B (Table 2), and using the minimum and maximum added 
flux of the secondary star, estimate the limits of their Binary HZs using equation (2). 

Table 3 shows the estimates of the boundaries of the Binary HZ around each star.
The terms Max and Min in this table correspond to the planet-binary configurations of $(\theta,\nu)=(0,0)$ and $(0,{180^\circ})$
(see figure 1 for the definition of $\theta$ and $\nu$) where the planet receives the maximum and minimum total flux
from the secondary star, respectively. As shown here, each star of the $\alpha$ Centauri system  
contributes to increasing the limits of the Binary HZ around the other star. 
Although these contributions are small, they extend the estimated limits of the Binary HZ by a noticeable amount. 
This is primarily due to the high luminosity of $\alpha$  Cen A and the relatively large eccentricity of the binary which 
brings the two stars as close as $\sim 11.3$ AU from one another (and as such making planet formation very difficult 
around $\alpha$ Cen A).

Generalizing our study, we examined the effect of increasing eccentricity in an $\alpha$ Cen-like system
(with similar G2V and K1V stars) by calculating the critical distance around both stars of this binary 
for which the entire HZ will be dynamically stable (i.e., ${l_{\rm out}}\leq {a_{\rm Max}}$). For the G2V star, 
the stability limit is given by

\begin{equation}
{a_{\rm Max}}\,=\,23.5 \Big(0.2892 \,-\, 0.361446\, {e_{\rm Bin}}\,+\, 0.05892\, {e_{\rm Bin}^2}\Big)
\end{equation}

\noindent
and for the K1V star, it is equal to

\begin{equation}
{a_{\rm Max}}\,=\,23.5 \Big(0.2584 \,-\, 0.313974\, {e_{\rm Bin}}\,+\, 0.042882\, {e_{\rm Bin}^2}\Big)
\end{equation}

\noindent
The maximum value of the binary eccentricity for which an Earth-like planet around the G2V star can have a 
stable orbit at the outer boundary of the narrow HZ is 0.62. For the empirical HZ, this maximum eccentricity
reduces to 0.59. For all values of the binary eccentricity smaller than 0.62 (0.59), the entire narrow 
(empirical) HZ around the G2V star will be stable. For the K1V star, the maximum binary eccentricity that allows 
the entire HZ to be stable is 0.73 for the narrow, and 0.71 for the empirical Binary HZs. 

Given the eccentricity of the $\alpha$ Cen binary $({e_{\rm Bin}}=0.518)$, both narrow 
and nominal HZs for $\alpha$ Cen A and B are stable. Figure 7 shows the maximum 
effective flux of a G2V star received by an Earth-like planet in the HZ of the K1V star 
during the secondary's (i.e., the G2V star) periastron passage. 
Note that this is a short-lived flux that can
be buffered by the planet's atmosphere and reduces the secondary's heating effect in its closest approach. 
The maximum flux of the secondary is only used here to estimate the maximum shift from the single star HZ to the Binary HZ.

In the $\alpha$ Cen system, where the binary eccentricity is 0.518, the stability limit around the
primary G2V star is at $\sim 2.768$ AU. This limit is slightly exterior to the outer boundary of the star's narrow and
empirical HZs. Although the latter suggests that the HZ of $\alpha$ Cen A is 
dynamically stable, the close proximity of this region to the stability limit may have strong consequences on the actual formation 
of an Earth-like planet in this region (see e.g. Th\'ebault et al. 2008, Eggl 2012).

\subsection {Binary HZ - Example: HD 196885}
HD 196885 is a close main sequence S-type binary system with a semimajor axis of 21 AU and eccentricity of 
0.42 (Chauvin et al. 2011).  The primary of this system (HD 169885 A) is an F8V star with a $T_{\rm Star}$ of 6340 K,  
mass of 1.33 ${M_{\rm Sun}}$, and luminosity of 2.4 $L_{\rm Sun}$. The secondary star (HD 196885 B) is an M1V 
dwarf with a mass of $0.45 {M_{\rm Sun}}$. Using the mass-luminosity relation $L \sim {M^{3.5}}$ 
where $L$ and $M$ are in solar units, the luminosity of this star is approximately $0.06 {L_{\rm Sun}}$,
and we consider its effective temperature to be ${T_{\rm Star}}=3700$ K.
The primary of HD 196885 hosts a Jovian-type planet suggesting that the mass-ratio and orbital elements of this binary 
allow planet formation to proceed successfully around its primary star. We assume that terrestrial planet formation 
can also successfully proceed around both stars of this binary and can result in the formation of Earth-sized objects. 

To estimate the boundaries of the Binary HZ of this eccentric system, we ignore its known giant planet and use
equation (2) considering a fictitious Earth-like planet in the HZ.  We calculate the spectral weight factor 
$W(f,T)$ for both stars of this system (Table 2) and estimate the locations of the inner and outer 
boundaries of the binary's HZ (Table 3). 

As expected (because of the large periastron distance of the binary, 
and the secondary star being a cool M dwarf), even the maximum flux from the secondary star does not have a 
noticeable contribution to the location of the HZ around the F8V primary of HD 196886. However, being a 
luminous F star, the primary shows a small effect on the location of the HZ around the M1V secondary star (Table 3).

Generalizing our study, we examined the effect of increasing eccentricity in an HD 196885-like system 
(with similar F8V and M1V stars) by calculating the critical distance around both stars of this binary for 
which the entire HZ will be dynamically stable (i.e., ${l_{\rm out}}\leq {a_{\rm Max}}$). Around the F8V star the 
stability limit is given by

\begin{equation}
{a_{\rm Max}}\,=\,21\Big(0.36786\,-\, 0.482742\,{e_{\rm Bin}}\,+\, 0.1\, {e_{\rm Bin}^2}\Big)
\end{equation}

\noindent
Around the M1V secondary, the critical distance is at

\begin{equation}
{a_{\rm Max}}\,=\,21 \Big(0.18014 \,-\, 0.193258\, {e_{\rm Bin}}\,+\, 0.002\, {e_{\rm Bin}^2}\Big)
\end{equation}

\noindent
Using equation (7) for the case when the planet orbits the F8V star, the maximum value of $e_{\rm Bin}$
for which the Binary HZ will be stable is 0.59 for the narrow, and 0.57 for the empirical HZs. This suggests
that for all values of ${e_{\rm Bin}} \leq 0.57$, the HZ of the F8V star will be stable. Similar calculations
for the HZ around the M1V star indicate that the upper value of the binary eccentricity for which the
narrow and empirical HZs of the M1V star will be stable are ${e_{\rm Bin}} \leq 0.82$ and ${e_{\rm Bin}} \leq 0.81$, 
respectively. Figure 8 shows the maximum effective flux of each star of the binary received by an Earth-like planet 
in the HZ of the other star during their periastron passages. Similar to the case of
$\alpha$ Cen binary, this is a short-lived maximum flux that can be buffered by the planet's atmosphere and would 
reduce the secondary's heating effect at its closest distance. The maximum flux of the secondary is only used here 
to estimate the maximum shift of the HZ. In practice, the influence of a secondary in an S-type binary on the extent of 
the Binary HZ around the other component has to be determined in a case by case basis for the specific geometry of the system.

\section{Discussion: The Effect of the Planet Eccentricity}
We considered the (fictitious) Earth-like planet in the HZ to be in a circular orbit. 
In a close binary system, the gravitational 
effect of the secondary may deviate the motion of the planet from a circle and cause its orbit to become eccentric. 
Similar to the case of a planet orbiting a single star,
planet's eccentricity increases the annually averaged irradiation from the primary star by a factor 
of $(1-{e_{\rm Pl}^2})^{-1/2}$(see Williams \& Pollard 2002). 
In addition, for a planet in an eccentric S-type binary system, 
eccentricity can also increase the temporary maximum flux received
from the secondary star due to the decrease in planet-secondary distance. This effect, however, depends on the planet's
relative position to the secondary. We assume that one can average the secondary's flux on the planet over 
the orbital period of the planet, 
as a very conservative estimate on the secondary's influence. This estimate can be used for close binaries 
for which we assume that the planet's atmosphere can buffer the changing flux from the secondary, over the 
secondary's orbit. Detailed general circulation modeling (GCM) are needed to determine
the time, in terms of binary or planet orbital period, that the planet's atmosphere can efficiently
buffer the changing flux of the secondary star.

\section{Concluding Remarks}
 
We have presented a methodology for calculating the spectral weight factors of the individual stars in a binary
system which can be used to determine the boundaries of the Binary HZ.
The foundation of our calculation is based on the fact that the response of the planet's atmosphere 
to the radiation from the star depends strongly on the star's spectral energy distribution as well as the planet's 
cloud fraction and atmospheric composition. For a given atmospheric composition, stars with different SEDs deposit 
different amount of energy, implying that the flux of a star received at the top of the planet's atmosphere has 
to be weighted according to the stars SED. This is especially important in binary systems where the two stars are 
of different spectral types. We derived a formula for the spectral weight factor that takes into account the spectral 
type of the star, as well as the models of the HZ around the Sun.

To demonstrate the maximum effect of a secondary, we calculated the single-star HZ around the primary 
and showed the contribution of the secondary in an F-F, M-M, and F-M binary for circular orbits as well as for eccentric binary orbits. We then demonstrate how to calculate the Binary HZ and use two known planet-hosting binaries $\alpha$ Cen AB and HD 196886 as demosntrations. We choose these two system because their stellar components are main-sequence stars 
and they present real examples of binaries in which stars have considerably different luminosities. We bracket the Binary 
HZ of these systems and studied the connection between their stability and the binary eccentricity. Our study indicates that the 
effect of the secondary star on the location and width of the Binary HZ is generally small in an S-type system. In systems where 
the secondary is more luminous, its effect can influence the extent of the HZ, especially for close eccentric binaries 
(barring stability requirements). 

Habitability of a system also requires the existence of a terrestrial-class planet in the HZ. In order 
for an S-type binary to form such a planet and maintain its long-term stability, its orbital elements have to 
satisfy stringent conditions. As can be seen from the currently known close, S-type systems, the semimajor axes 
of these binaries are within the range of 17-20 AU and their eccentricities are limited to 
low to moderate values. The perturbation of the secondary star in these systems may excite the dynamics of the 
Earth-like planet in the Binary HZ and increase its orbital eccentricity slightly. However, such induced orbital eccentricities 
will not have drastic effects on the planet's habitability.

\acknowledgments
Special Thanks to S. Kane for useful comments. L.K. acknowledges support from NAI and DFG funding ENP Ka 3142/1-1. N.H. acknowledges support from the NAI under Cooperative Agreement NNA09DA77 at the Institute for Astronomy, University of Hawaii, NASA EXOB grant NNX09AN05G, HST grant HST-GO-12548.06-A, and Alexander von Humboldt Foundation. Support for program 
HST-GO-12548.06-A was provided by NASA through a grant from the Space 
Telescope Science Institute, which is operated by the Association of Universities for Research in Astronomy, Incorporated, 
under NASA contract NAS5-26555.

\clearpage
\begin{figure}
\center
\vskip 0.4in
\includegraphics[scale=0.65]{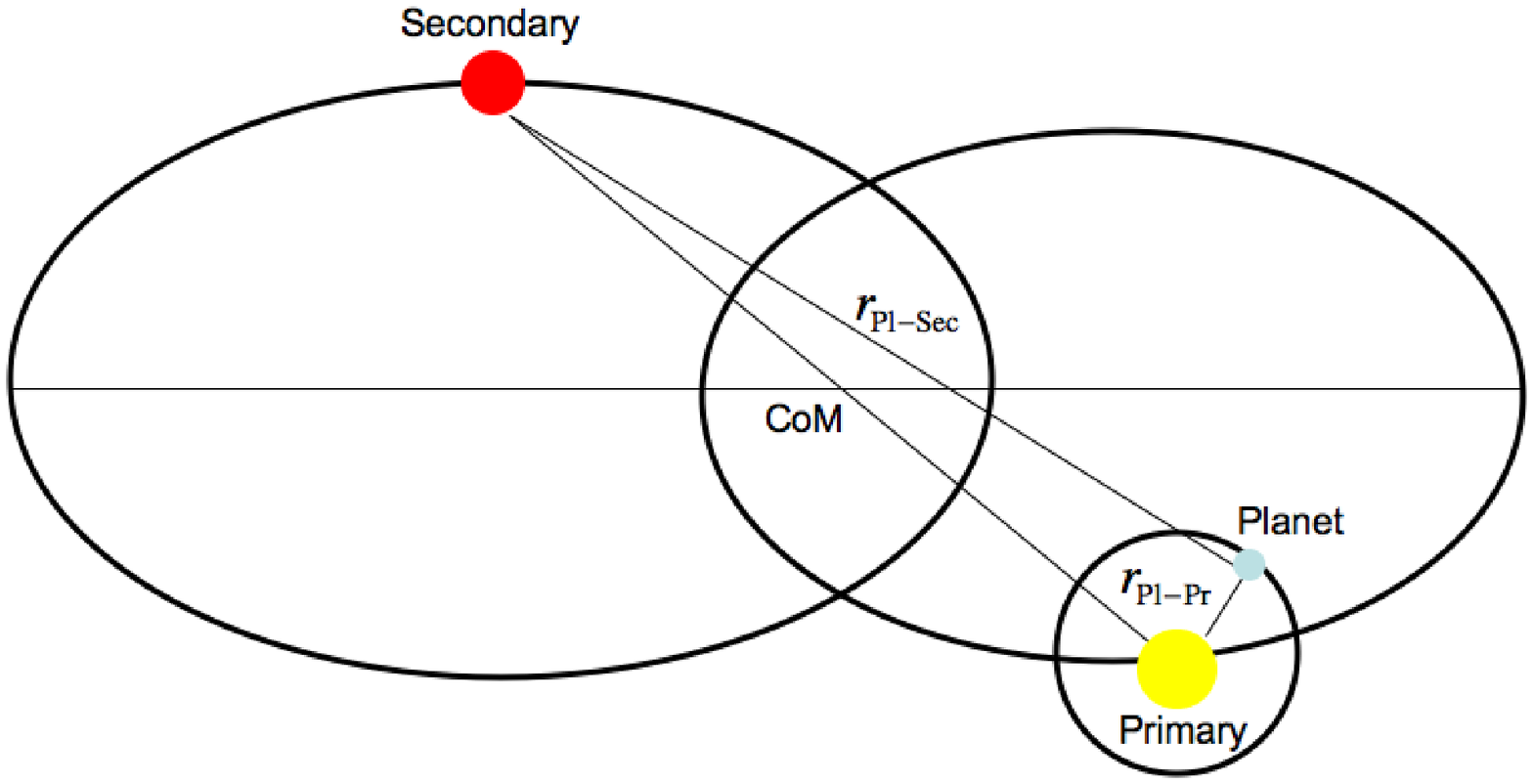}
\includegraphics[scale=0.65]{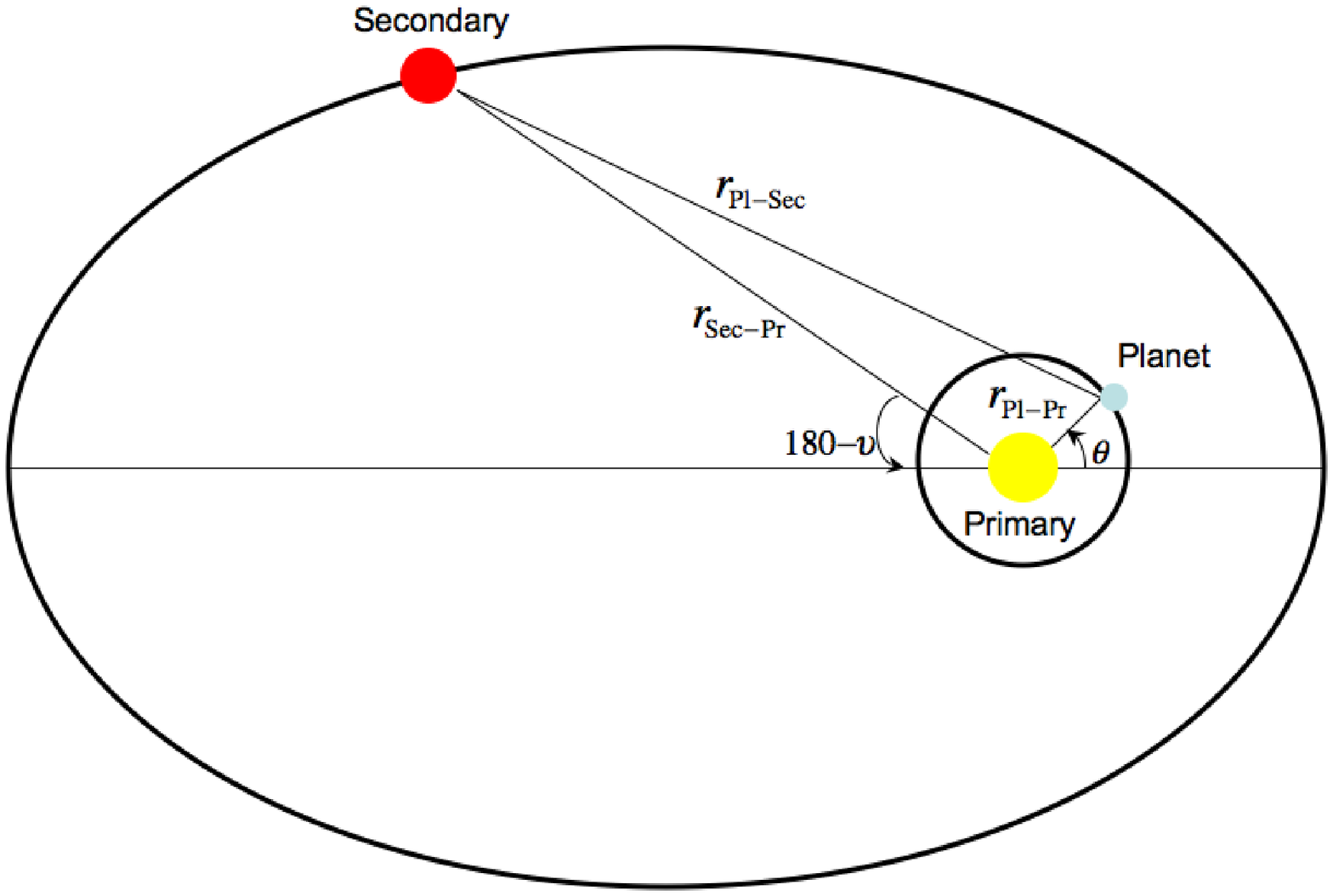}
\caption{A schematic presentation of an S-type system. The two stars of the binary, Primary and Secondary, 
revolve around their center of mass (CoM) while the planet orbits only one of the stars (top panel). 
It is, however, customary to neglect the motion of the binary around its CoM, and consider 
the motion of the secondary around a stationary primary (bottom panel).}
\end{figure}

\clearpage
\begin{figure}
\center
\vskip 1.2in
\includegraphics[scale=0.65]{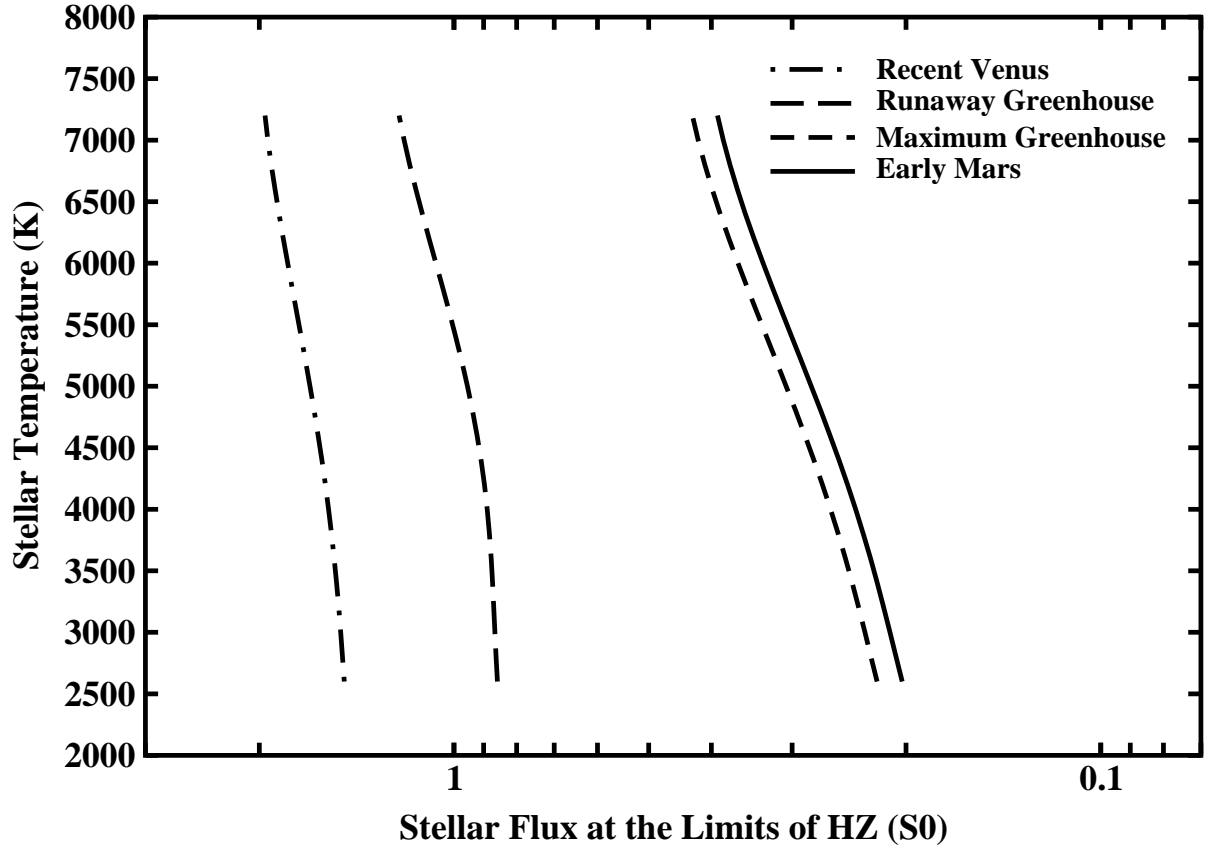}
\caption{Stellar flux at the top of an Earth-like planet's atmosphere, 
based on atmospheric models (Kopparapu et al. 2013a,b) when the planet is at the boundaries of the nominal 
and empirical HZs. The flux on the $x$-axis is scaled to the flux of the Sun at Earth's orbit $(S_0)$.}
\end{figure}

\clearpage
\begin{figure}
\center
\vskip 0.4in
\includegraphics[scale=1.]{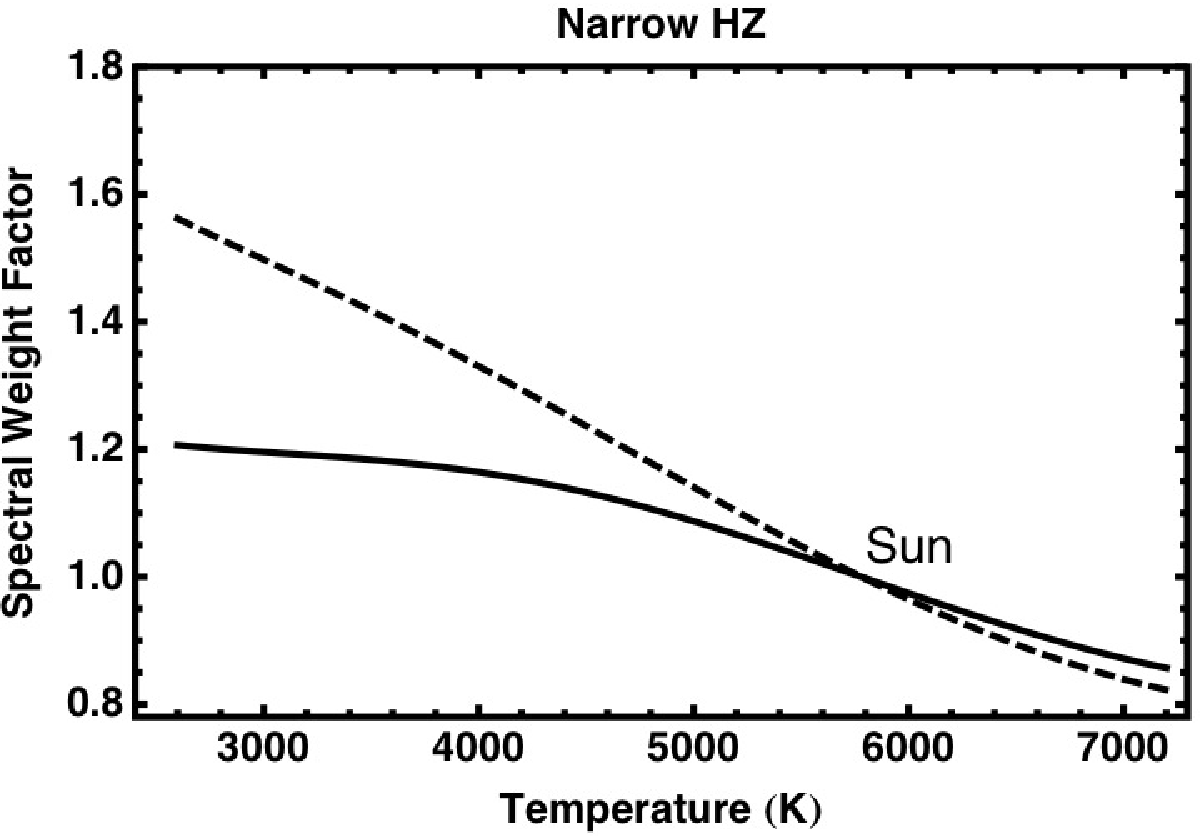}
\vskip 30pt
\includegraphics[scale=1.]{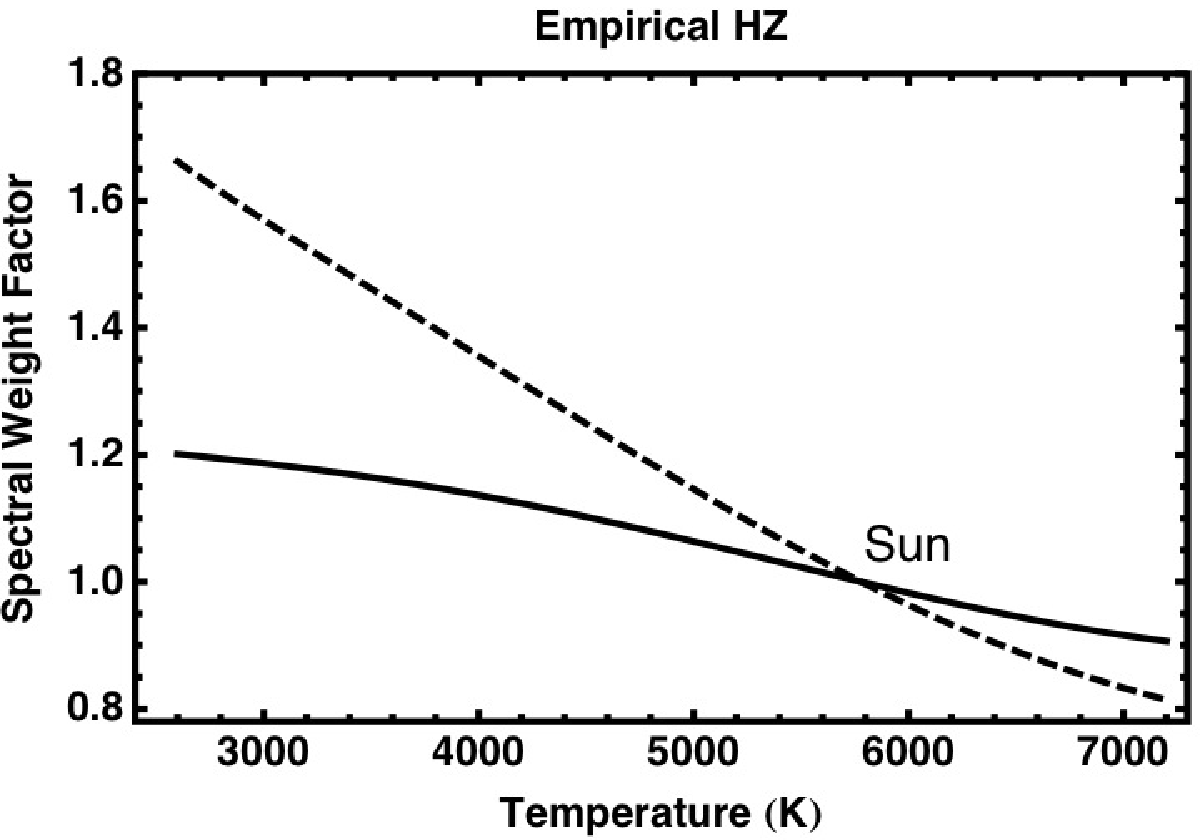}
\caption{Spectral weight factor $W(f,T)$ as a function of stellar effective temperature for the narrow (top) and 
empirical (bottom) HZs. The solid line corresponds to the inner and the dashed line corresponds to the outer
boundaries of HZ. We have normalized $W(f,T)$ to its solar value, indicated on the graphs (Sun).}
\end{figure}

\clearpage
\begin{figure}
\vskip 1.2in
\center
\includegraphics[scale=0.6]{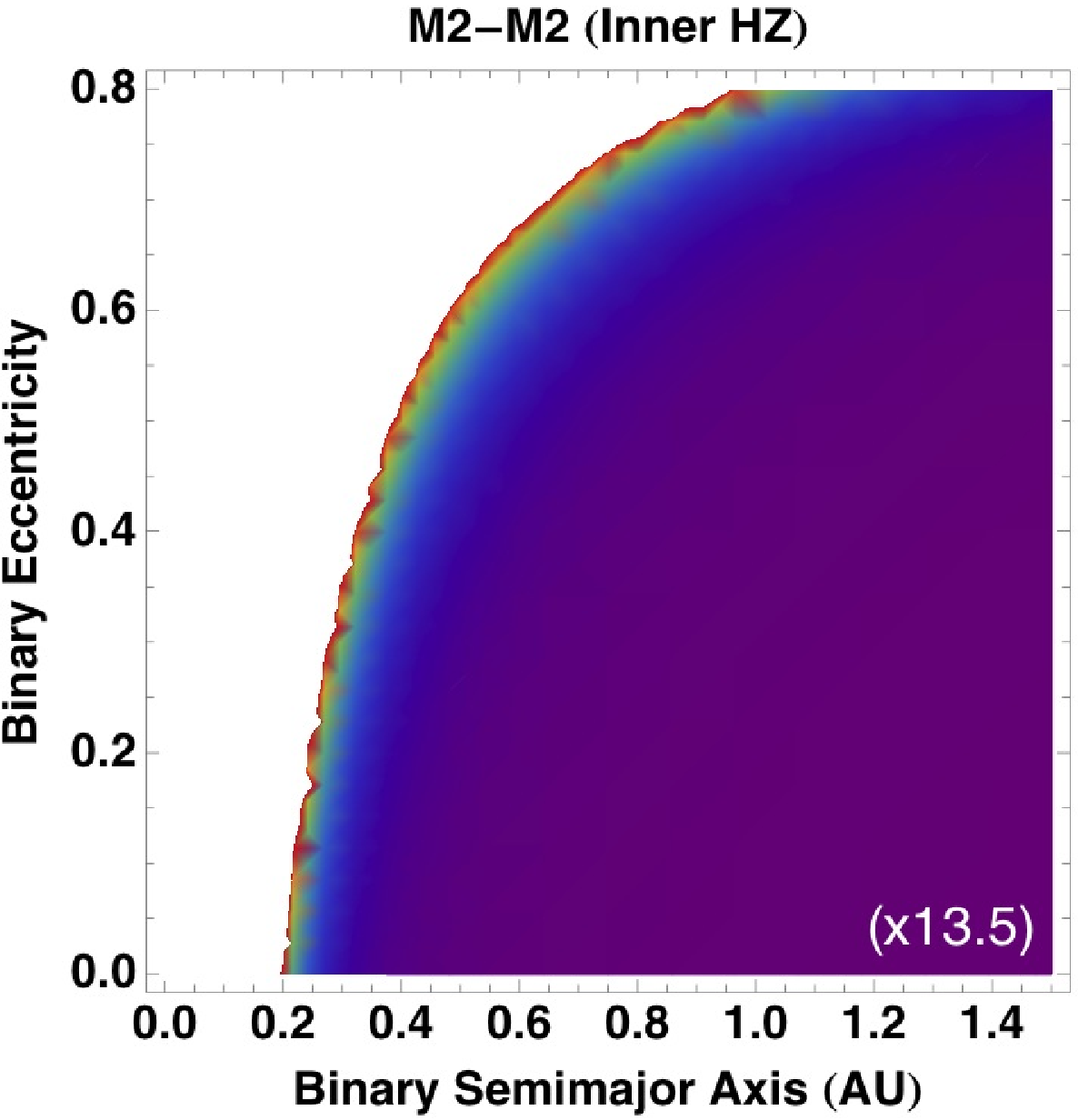}
\includegraphics[scale=0.6]{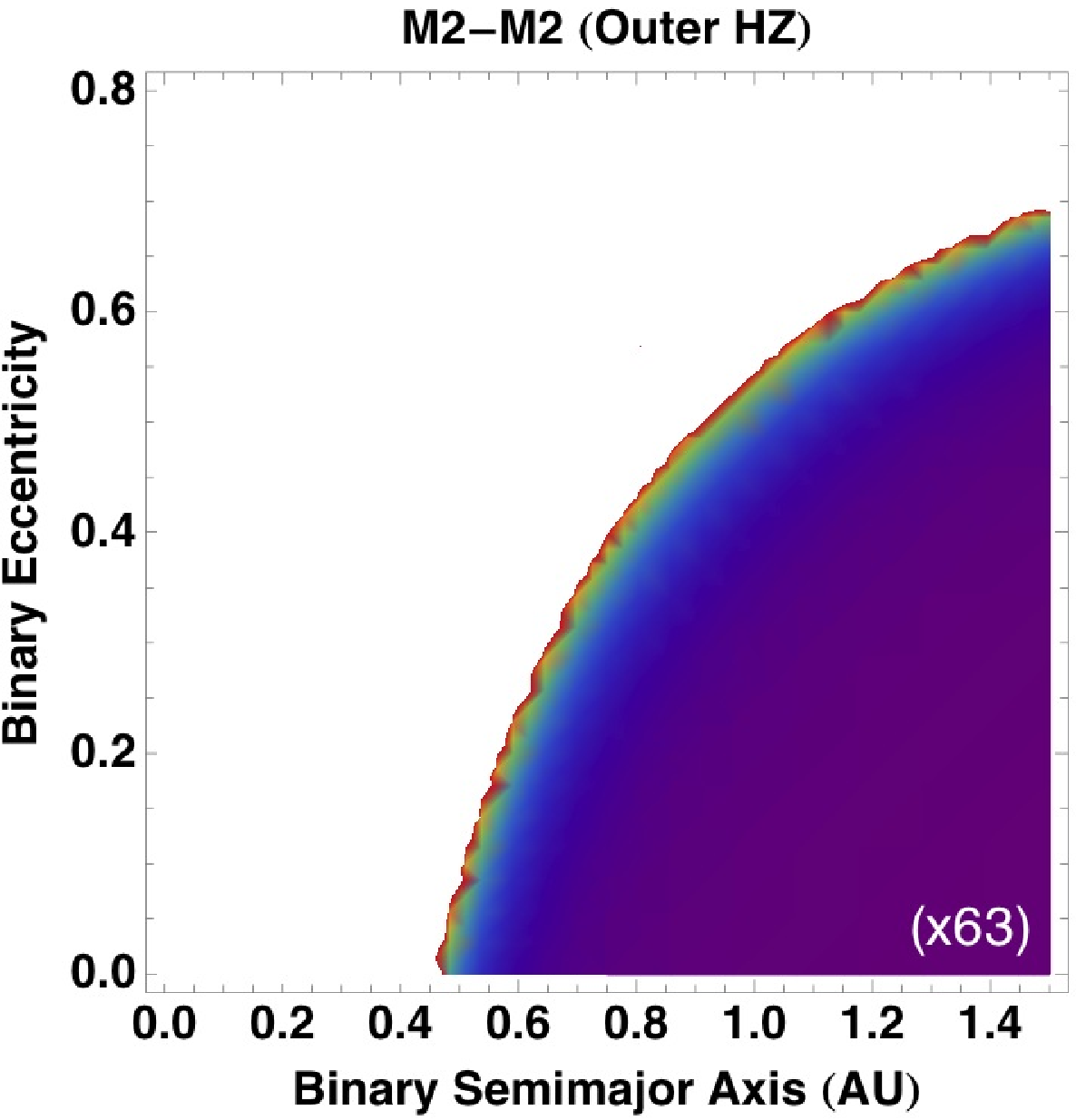}
\vskip 20pt
\includegraphics[scale=0.6]{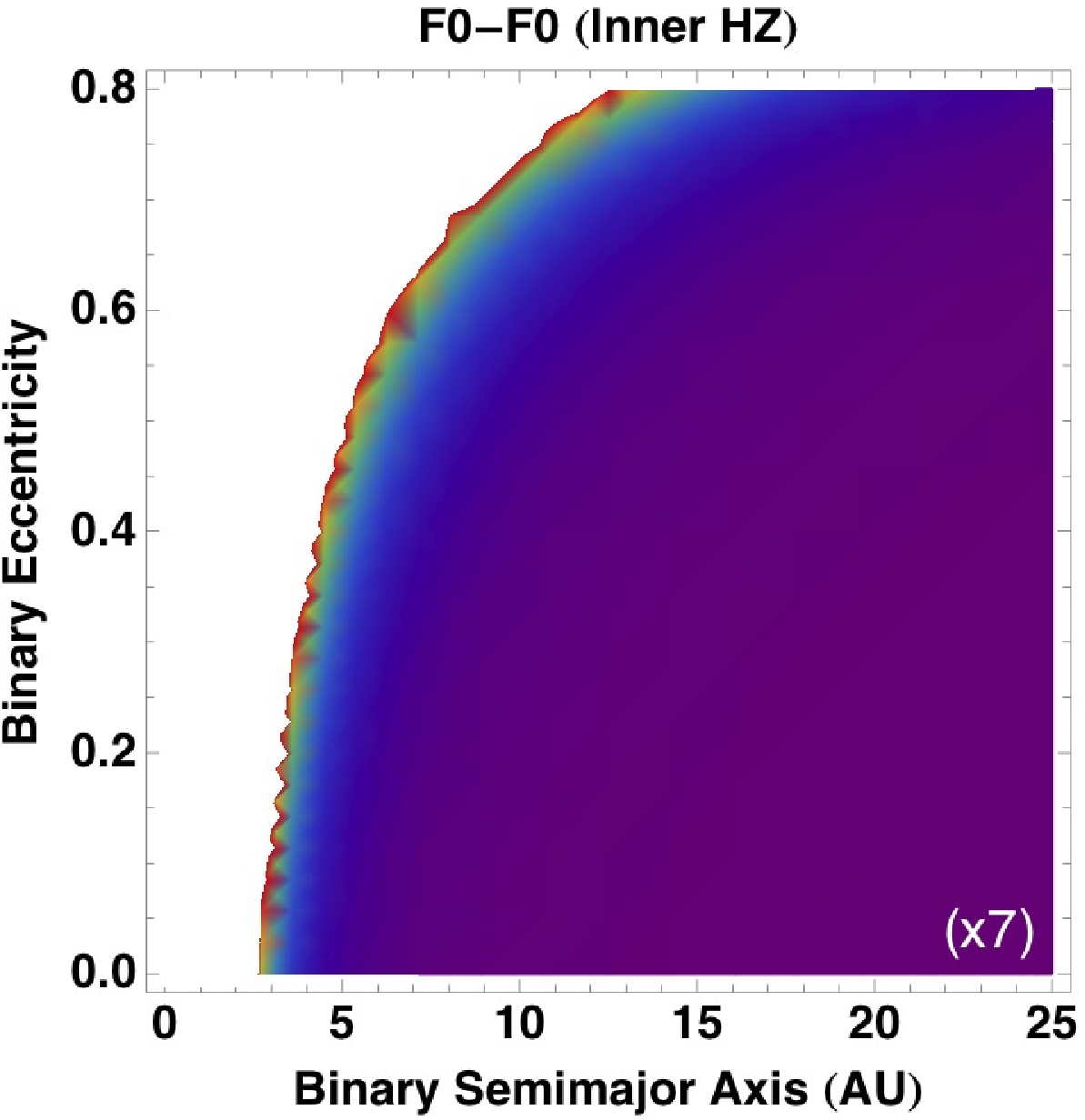}
\includegraphics[scale=0.6]{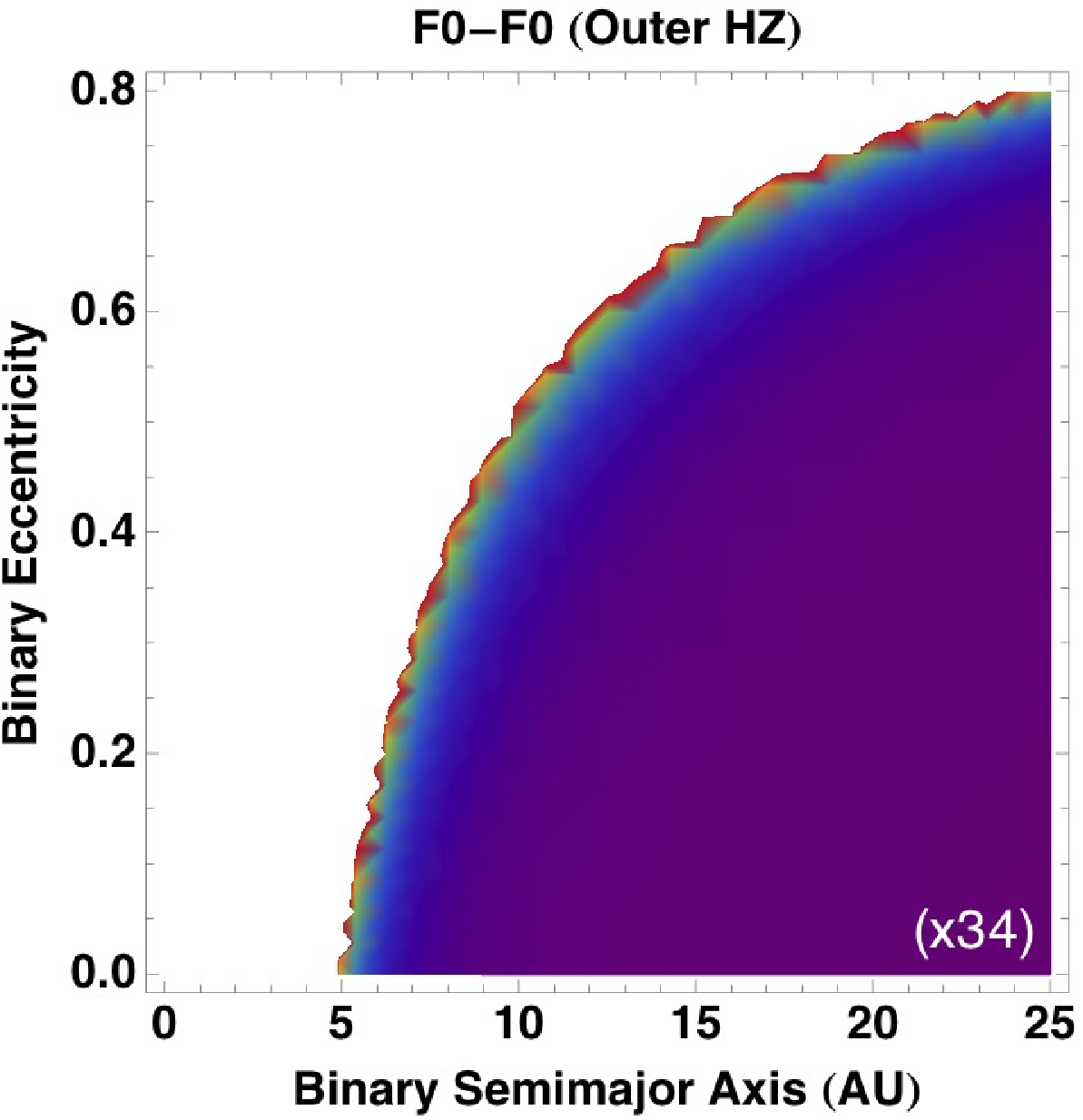}
\vskip 5pt
\includegraphics[scale=1]{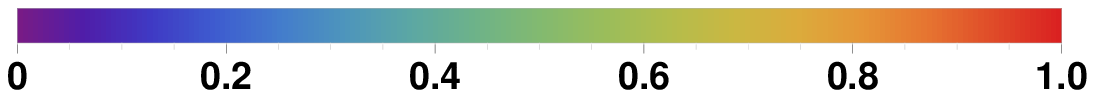}
\caption{Maximum contribution of the secondary star to the total flux received at the boundary of 
the single star HZ of the primary of an M2-M2 (top) and an F0-F0 (bottom) S-type binary. 
The color-coding, times the number on the lower right corner on each panel, represents the flux contributed by the 
secondary star at closest distance, relative to the flux received from the primary. Note that we do not considering stability of a fictitious planet here. The left (right) column corresponds to the inner (outer) limit of the empirical HZ.}
\end{figure}

\clearpage
\begin{figure}
\vskip 1.2in
\center
\includegraphics[scale=0.6]{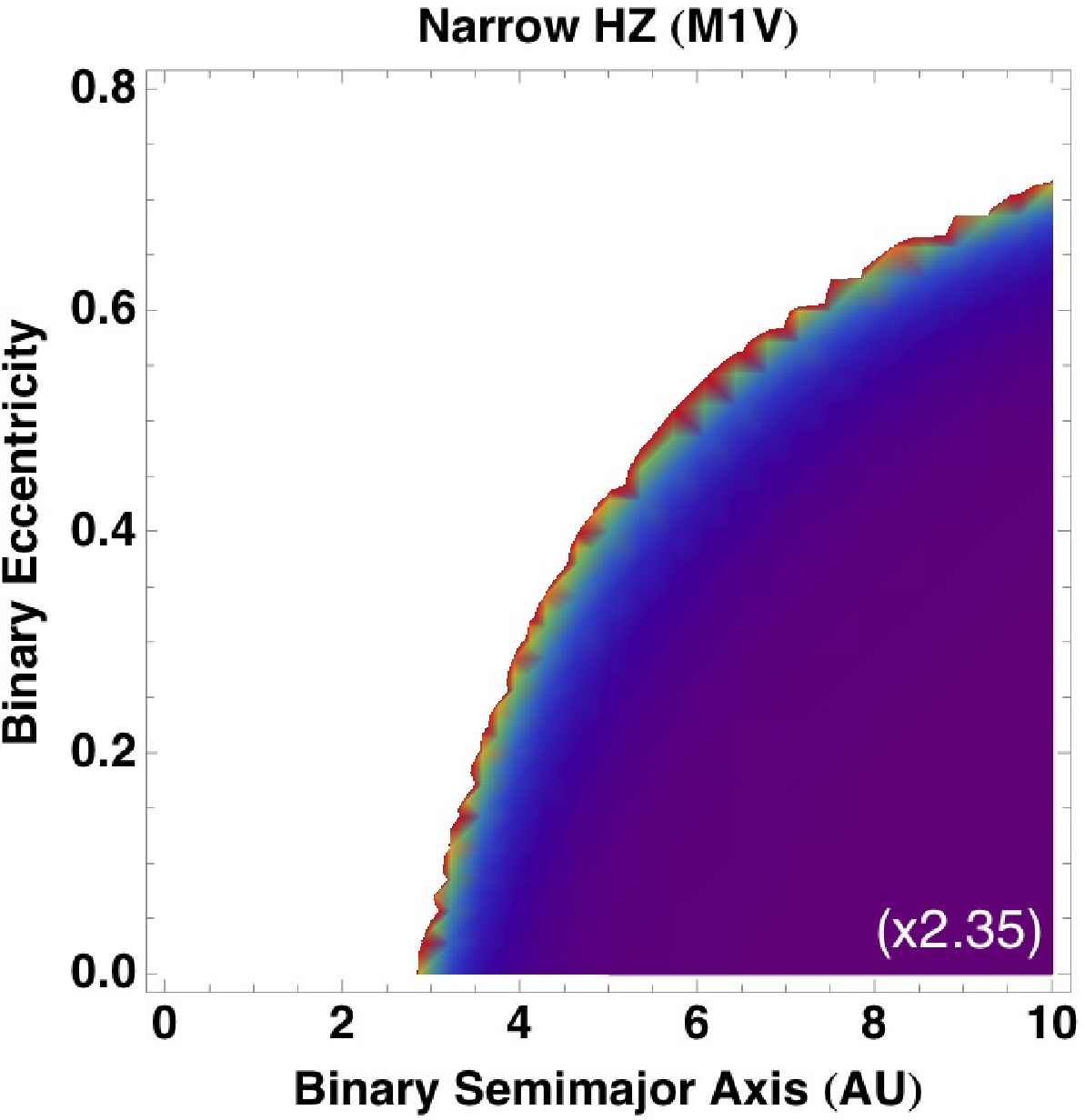}
\includegraphics[scale=0.6]{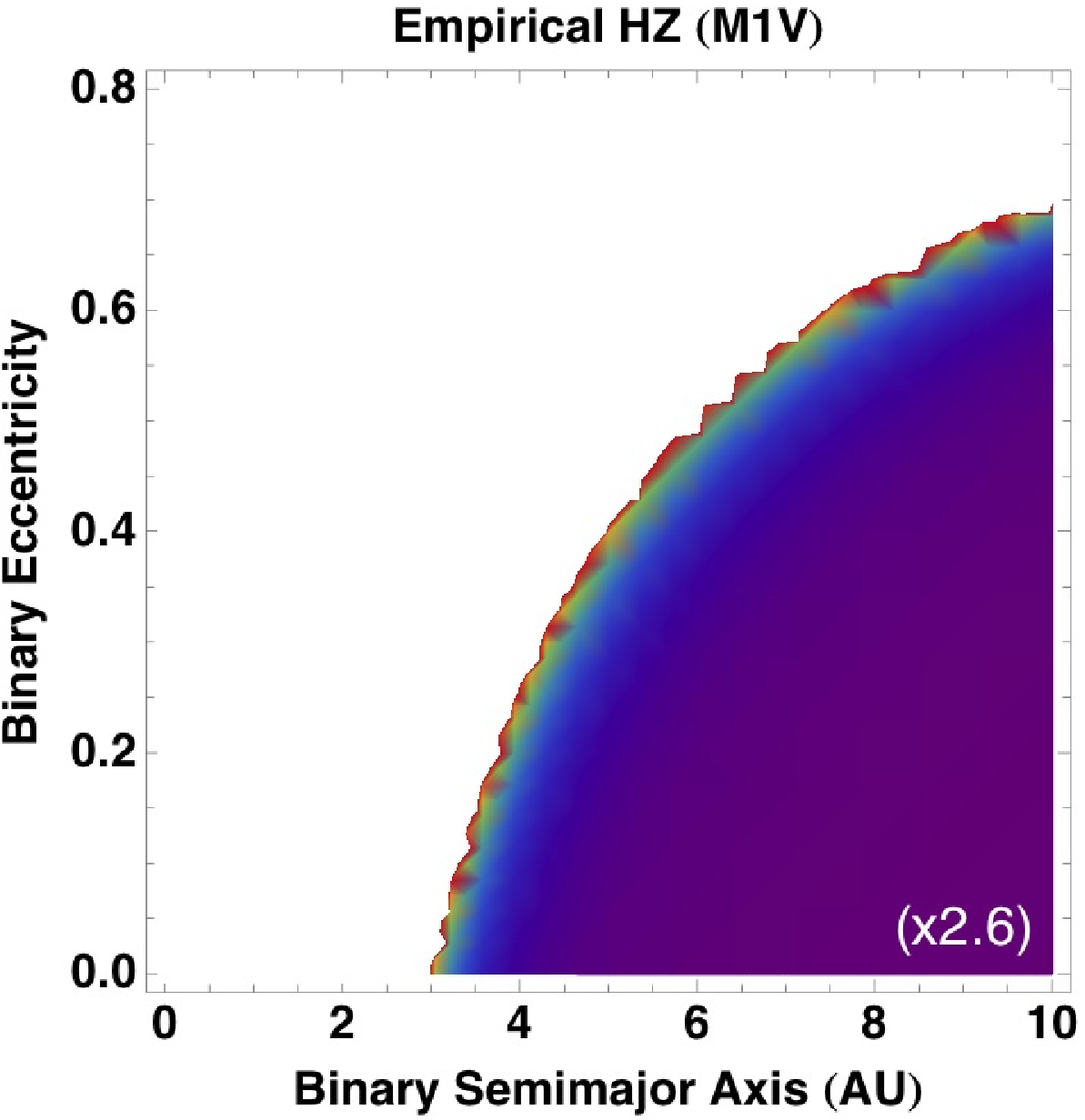}
\vskip 20pt
\includegraphics[scale=0.6]{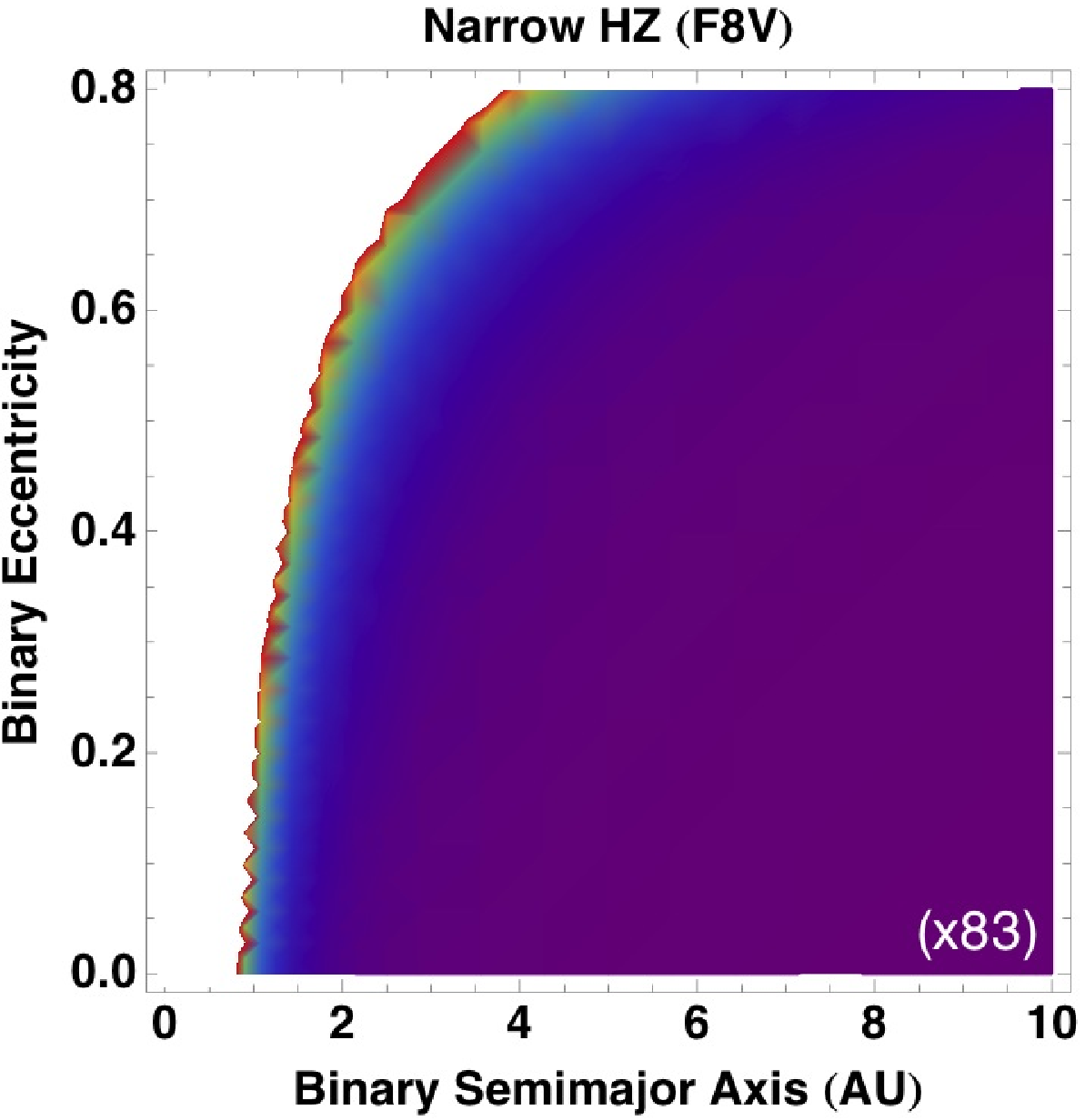}
\includegraphics[scale=0.6]{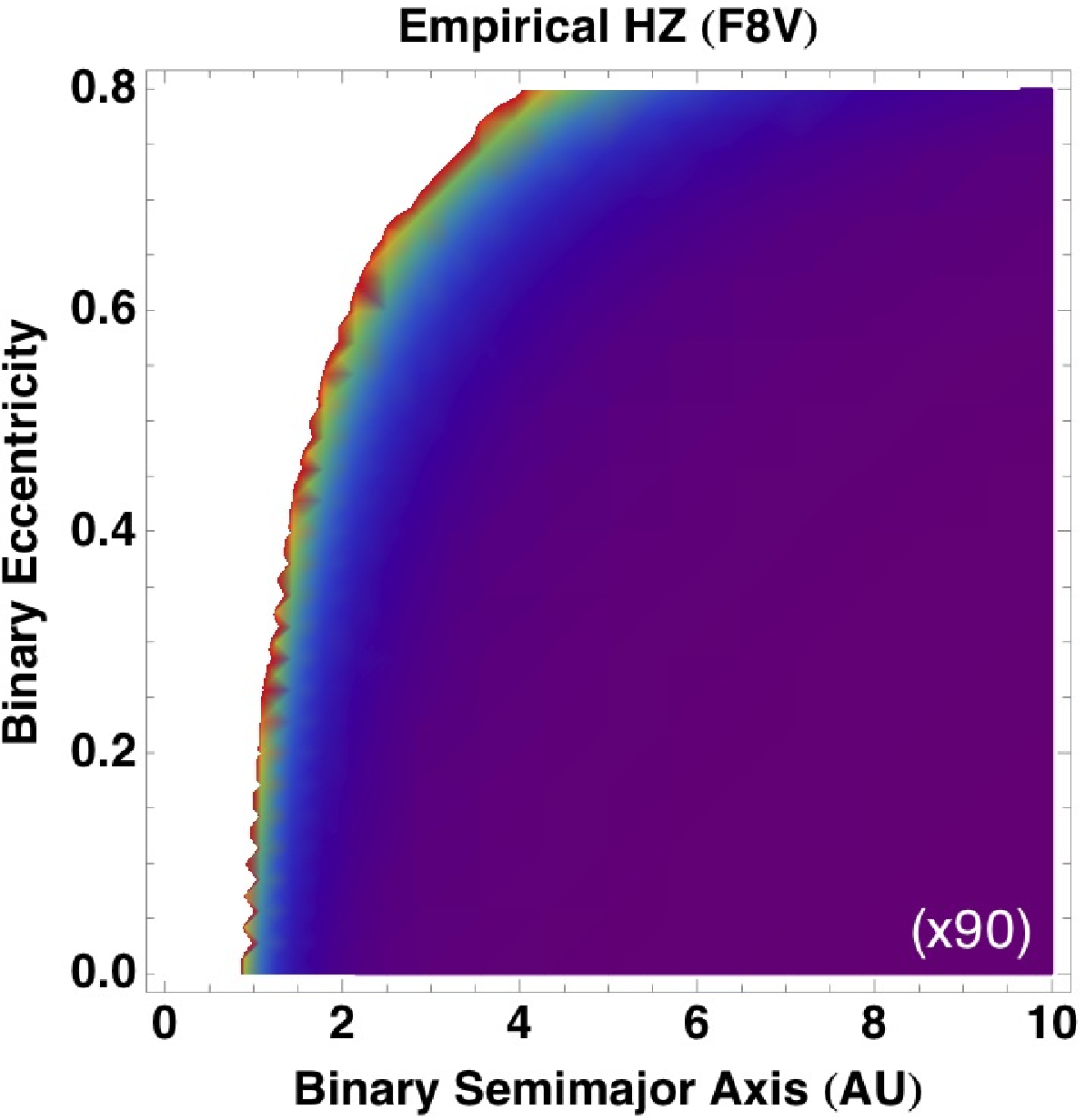}
\vskip 5pt
\includegraphics[scale=1]{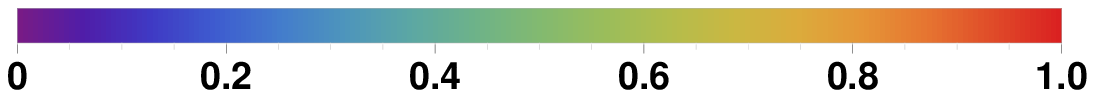}
\caption{Maximum contribution of the secondary star (top: M1V, bottom: F8V) to the total flux received at 
the outer limit of the single-star HZ of the primary star (top: F8V, bottom: M1V) in a F8V-M1V S-type binary system - similar to HD 196885. The color-coding, times the number on the lower right corner on each panel, represents the flux contributed by the 
secondary star at closest distance, relative to the flux received from the primary at the boundary of the primary's single star HZ. Note that we do not considering stability of a fictitious planet here. The left (right) column corresponds to the flux received at the outer limit of the narrow (empirical) HZ. }
\end{figure}

\clearpage
\begin{figure}
\center
\vskip 2in
\includegraphics[scale=0.5]{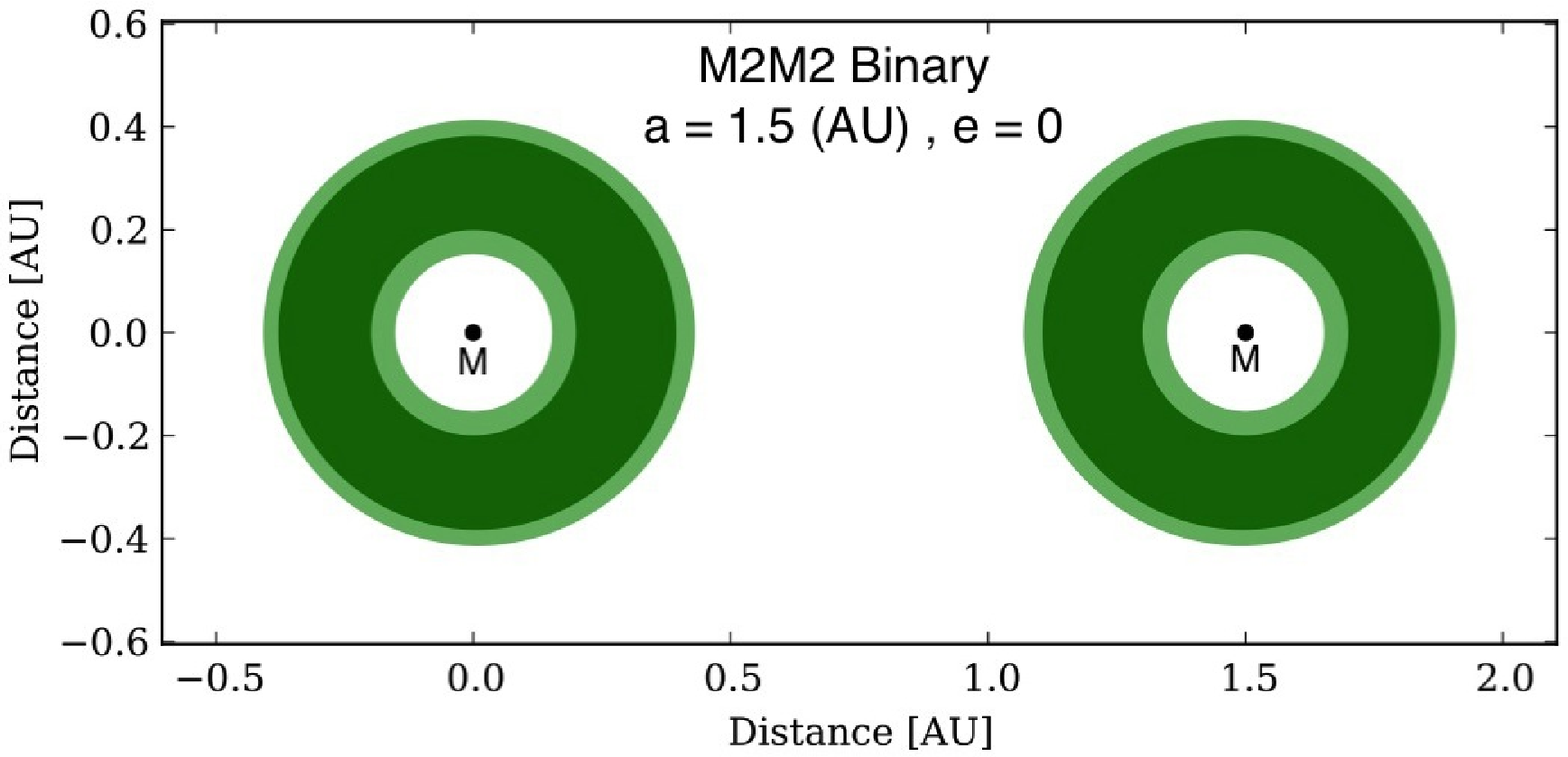}
\includegraphics[scale=0.5]{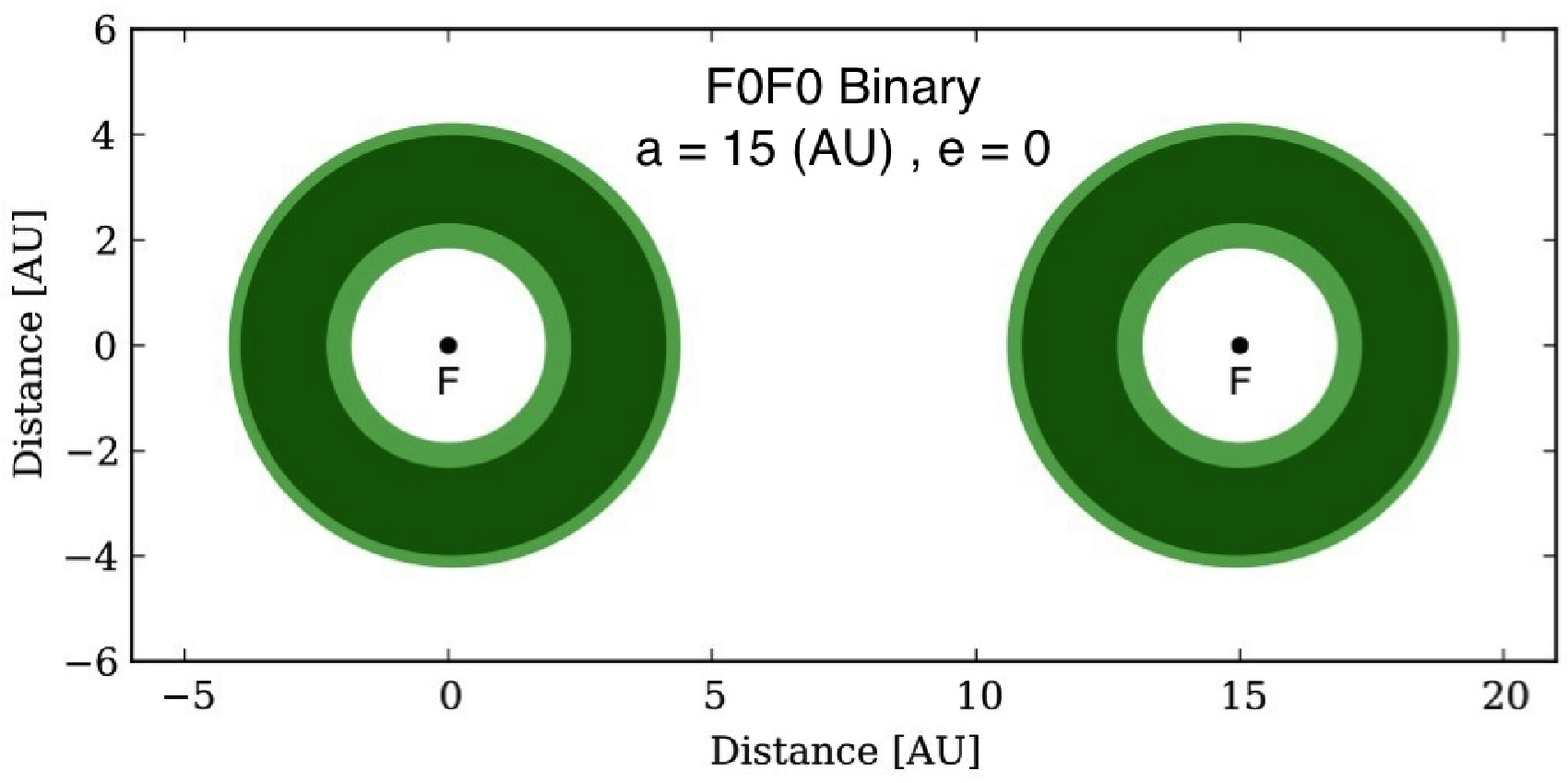}
\vskip 20pt
\includegraphics[scale=0.5]{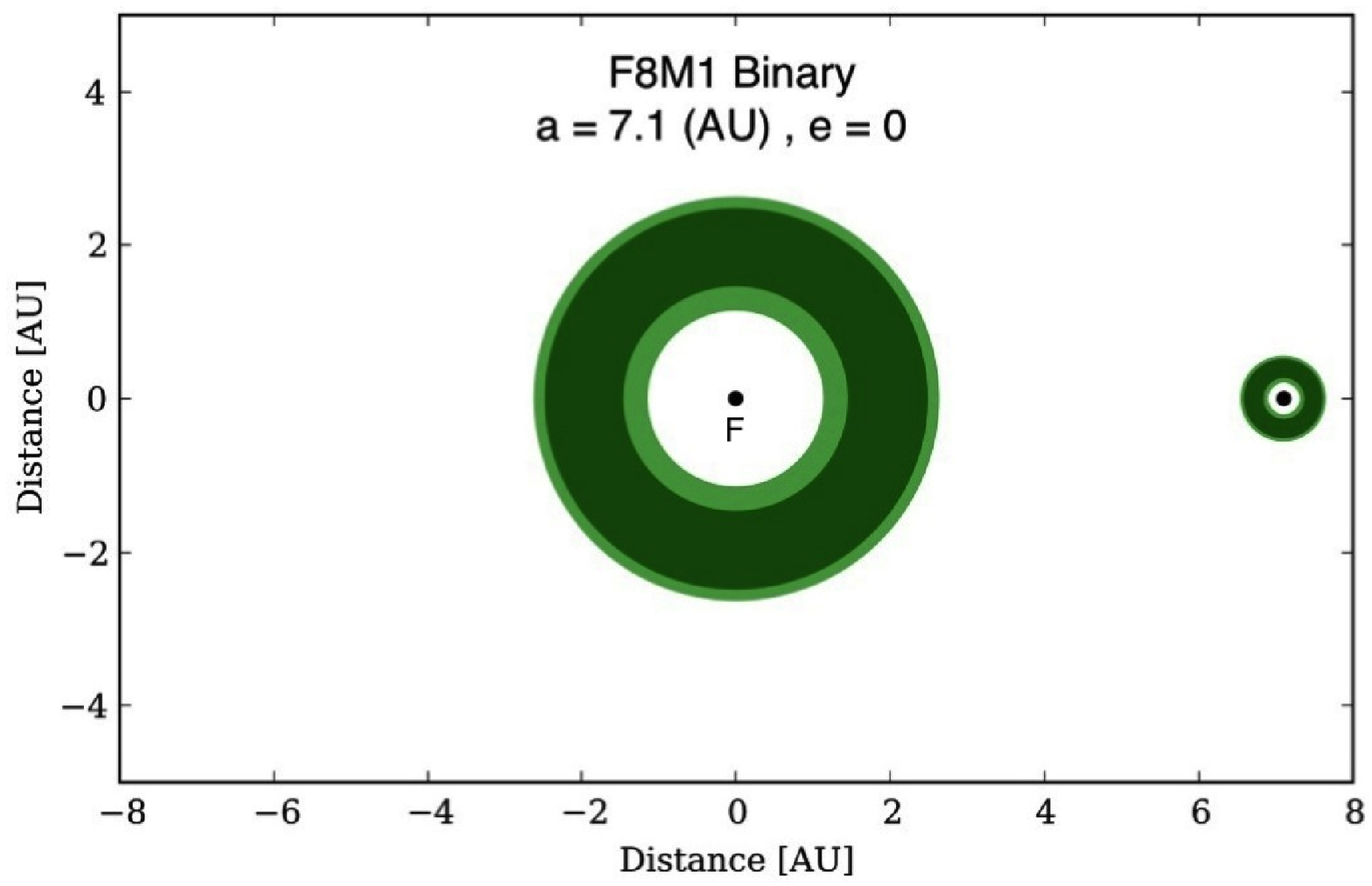}
\includegraphics[scale=0.5]{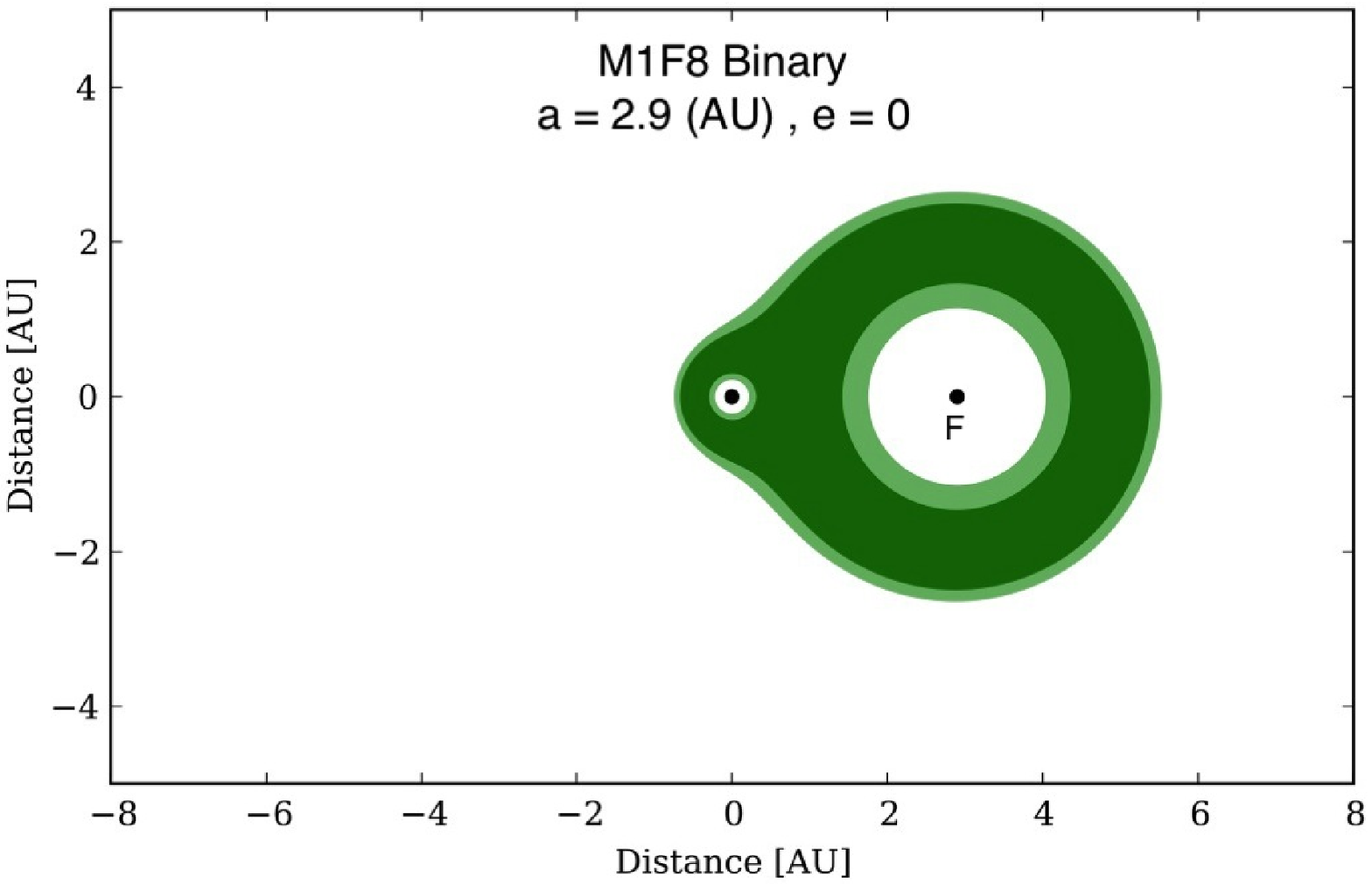}
\caption{Boundaries of the narrow (dark green) and empirical (light green) HZs in an M2-M2 (top left), F0-F0 (top right), 
and M1-F8 S-type binary star system (bottom two panels). Note that the primary is the star at (0,0). The
primary star in the bottom panels is the F8 star (left) and the M1 star (right). The semimajor axis of the binary has been
chosen to be the minimum value that allows the region out to the outer edge of the primary's empirical HZ to be stabile 
in a circular binary.}
\end{figure}

\clearpage
\begin{figure}
\center
\vskip 1.2in
\includegraphics[scale=0.5]{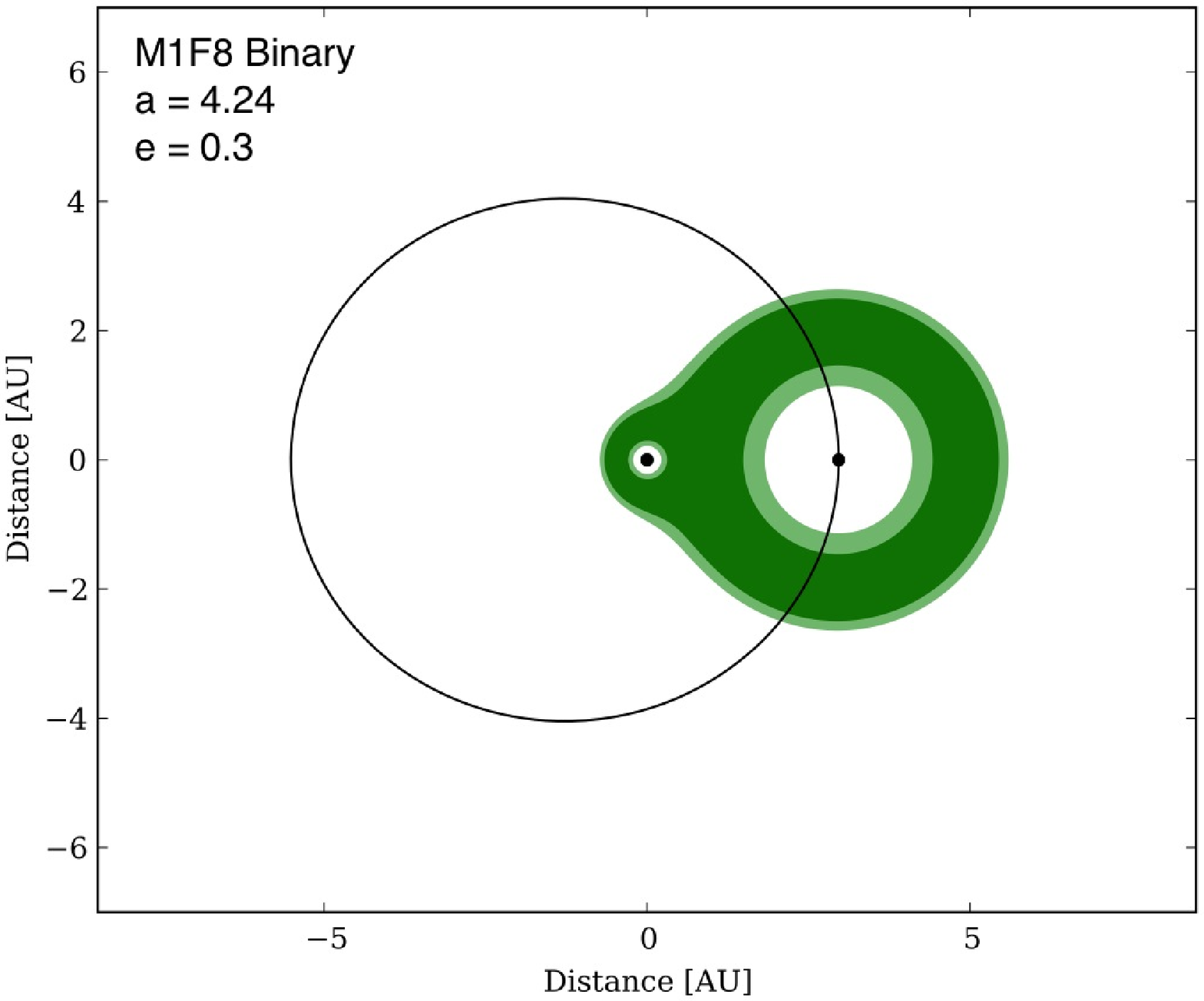}
\includegraphics[scale=0.5]{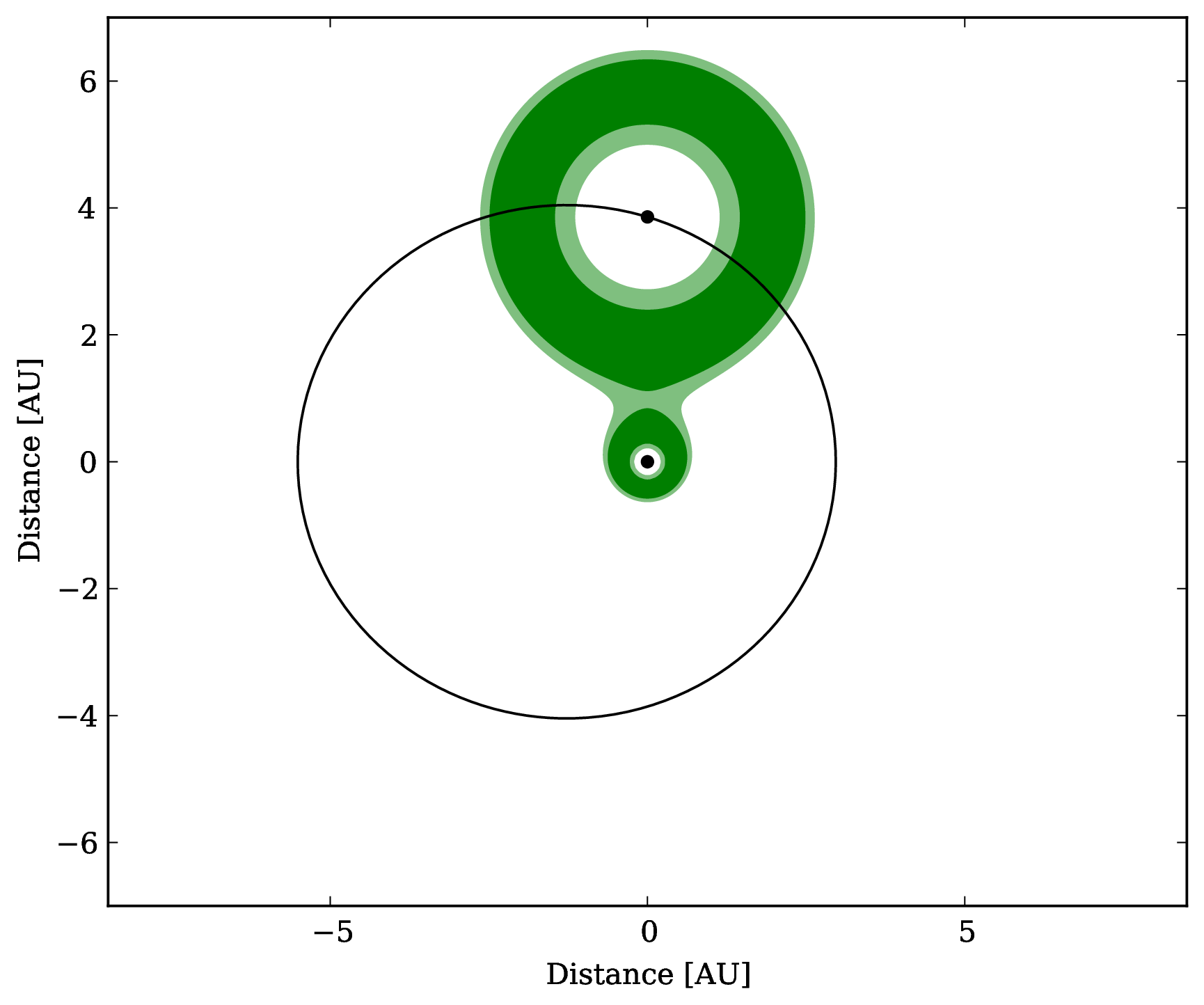}
\vskip 2.5in
\vskip 20pt
\includegraphics[scale=0.5]{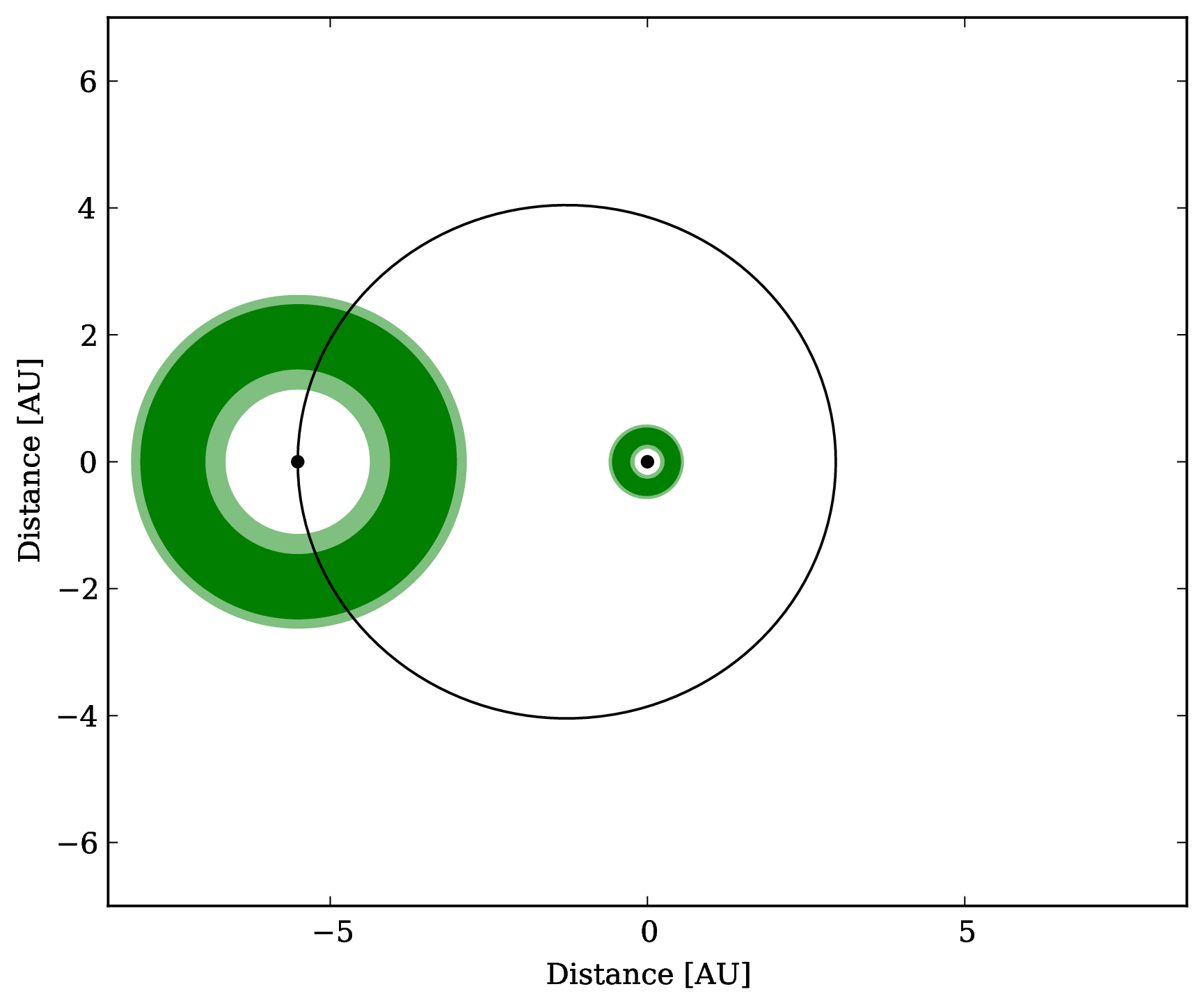}
\includegraphics[scale=0.5]{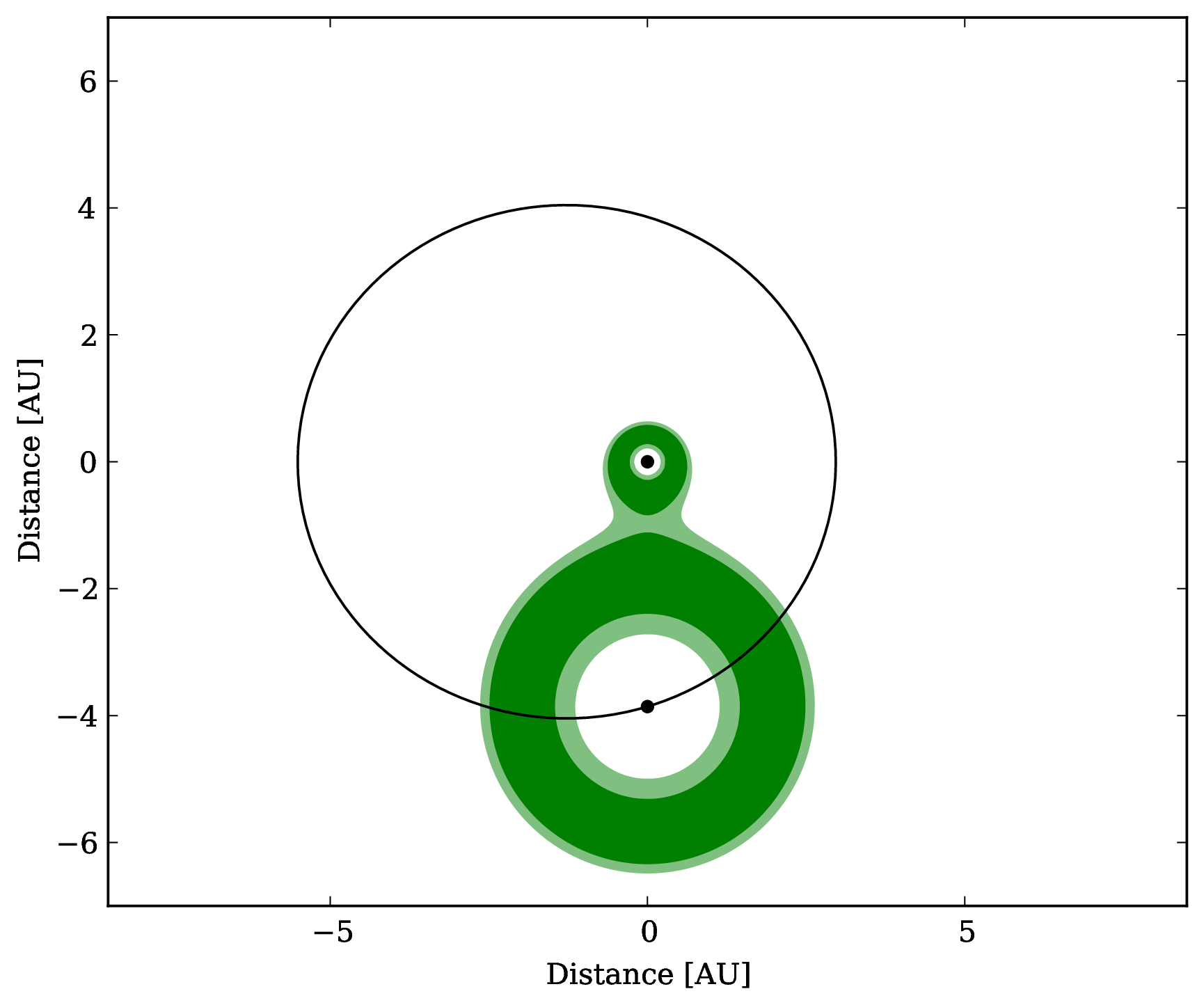}
\caption{Boundaries of the narrow (dark green) and empirical (light green) HZs in an M1-F8 binary. Note that the 
primary is the M1 star at (0,0). The panels show the effect of the F8 star while orbiting the primary starting from the
top left panel when the secondary is at the binary periastron. The semimajor axis of the binary has been
chosen to be the minimum value that allows the region out to the outer edge of the primary's empirical single-star HZ to be stable for a binary eccentricity of 0.3.}
\end{figure}

\clearpage
\begin{figure}
\center
\vskip 1.2in
\includegraphics[scale=0.5]{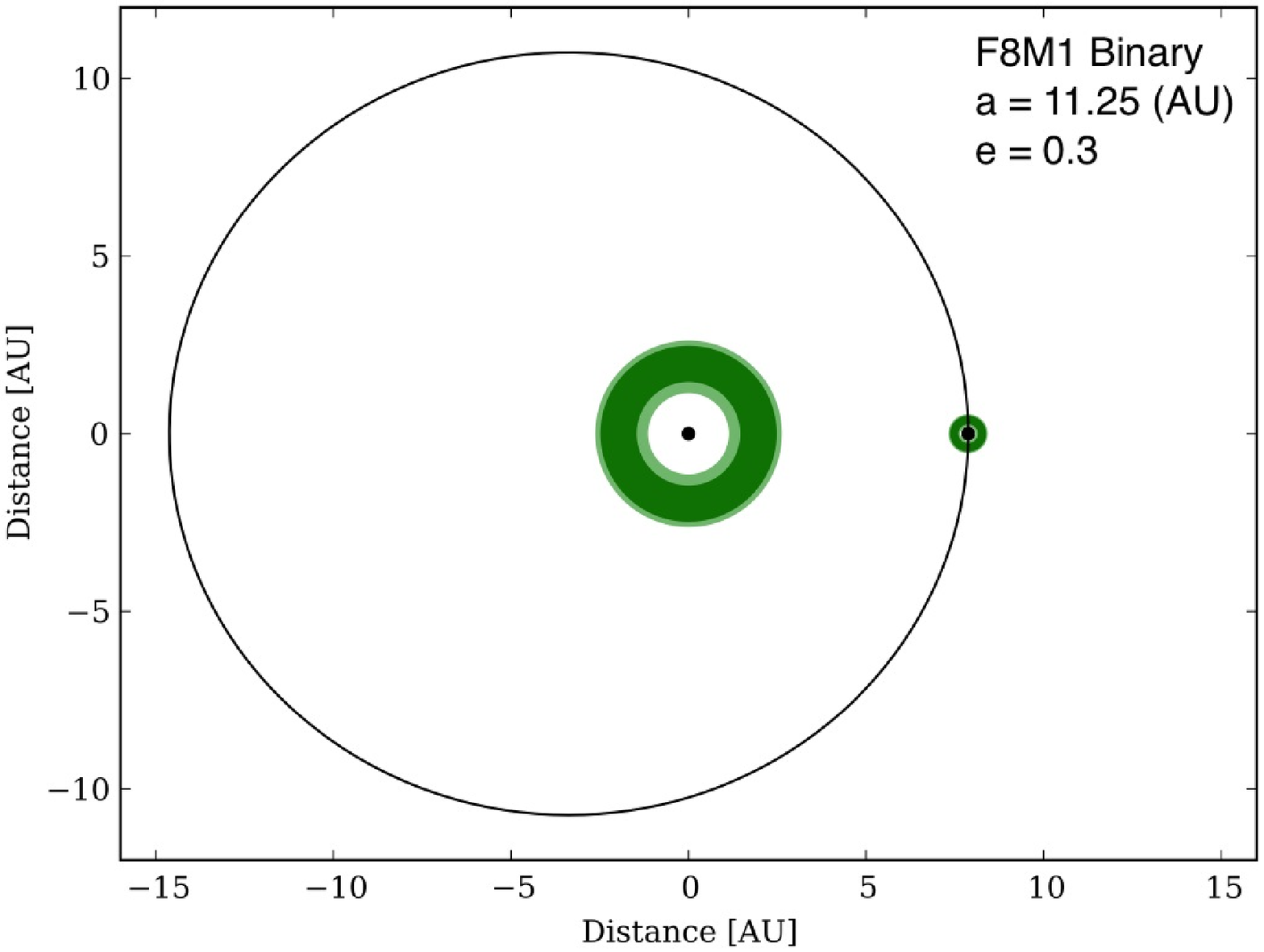}
\includegraphics[scale=0.5]{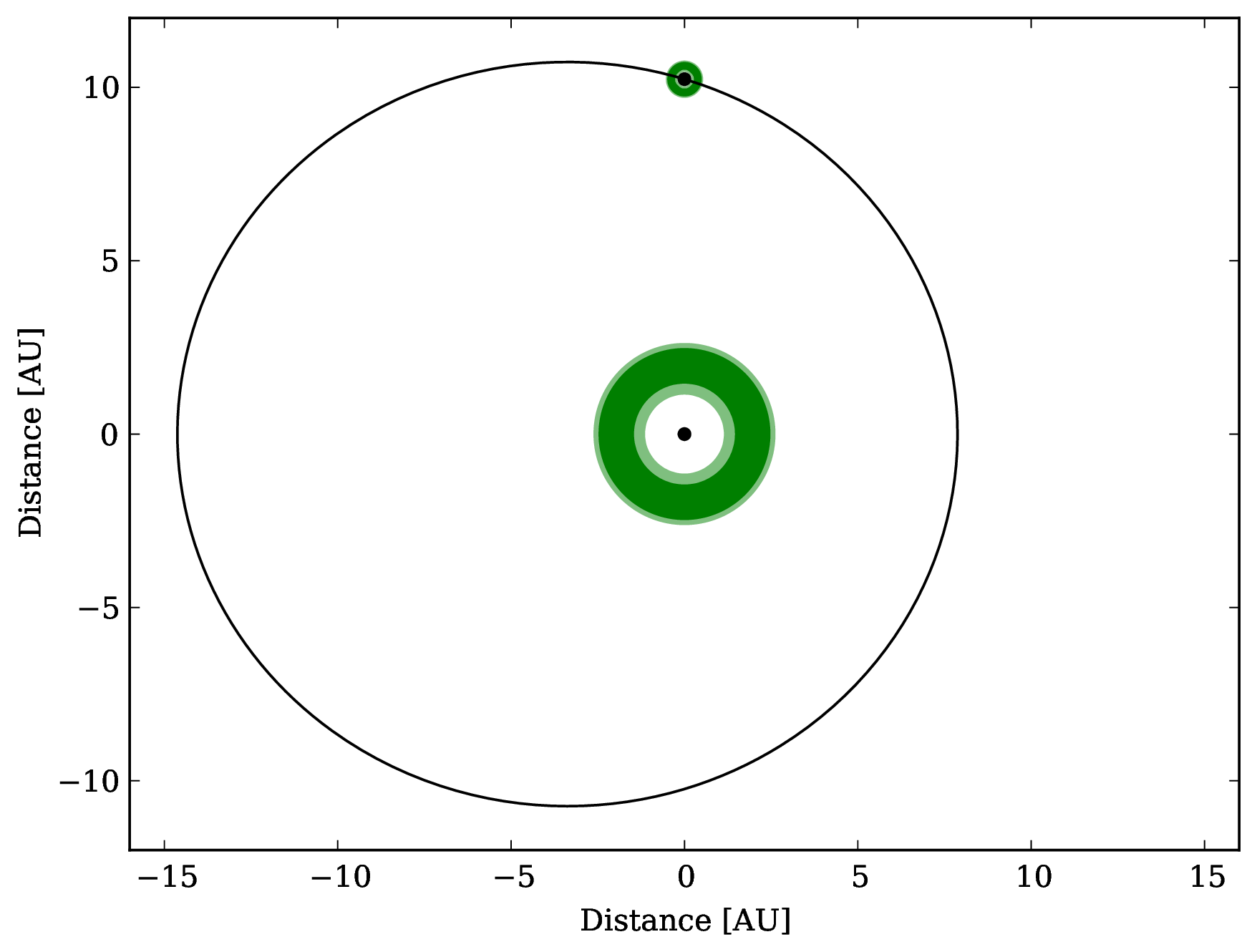}
\vskip 2.5in
\includegraphics[scale=0.5]{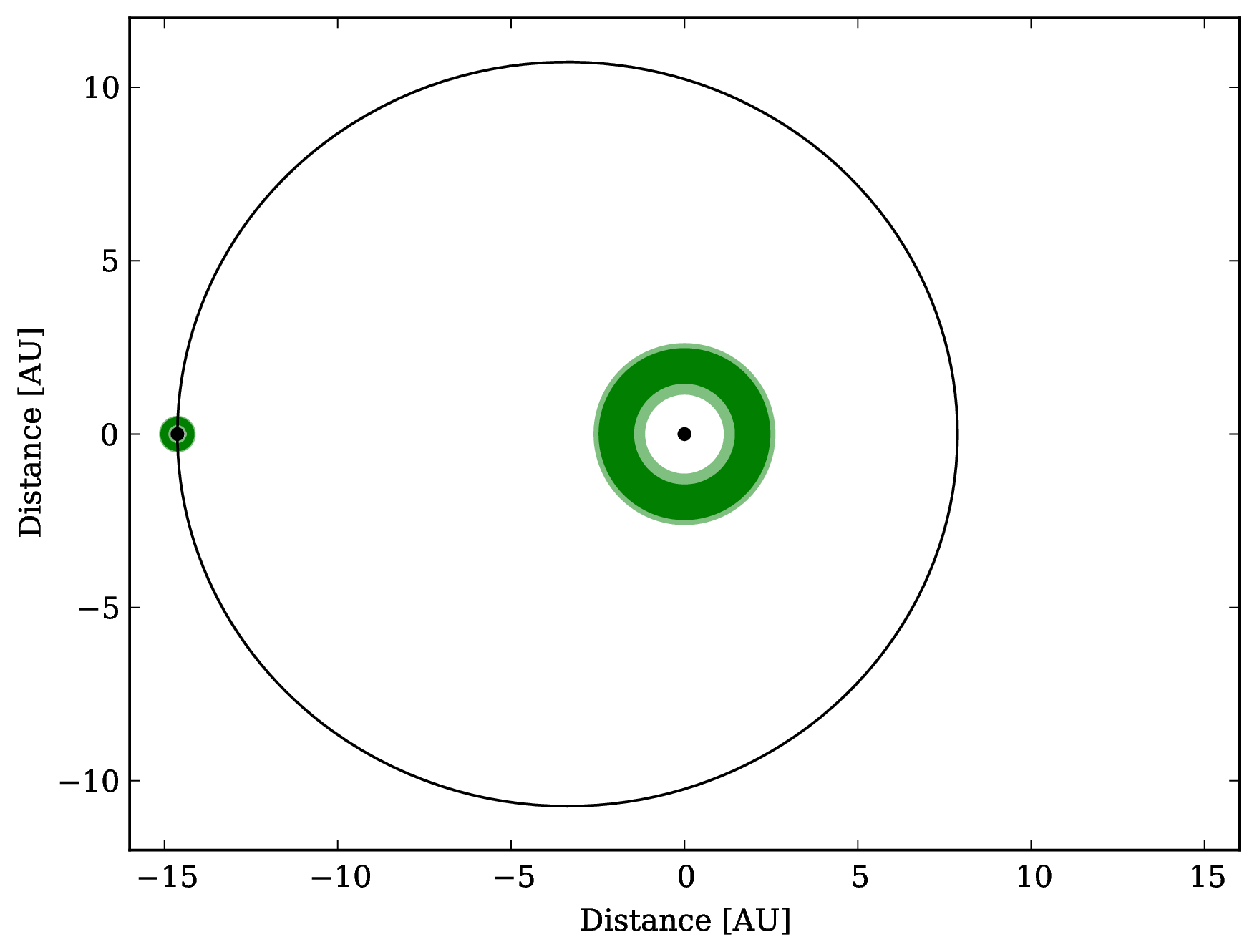}
\includegraphics[scale=0.5]{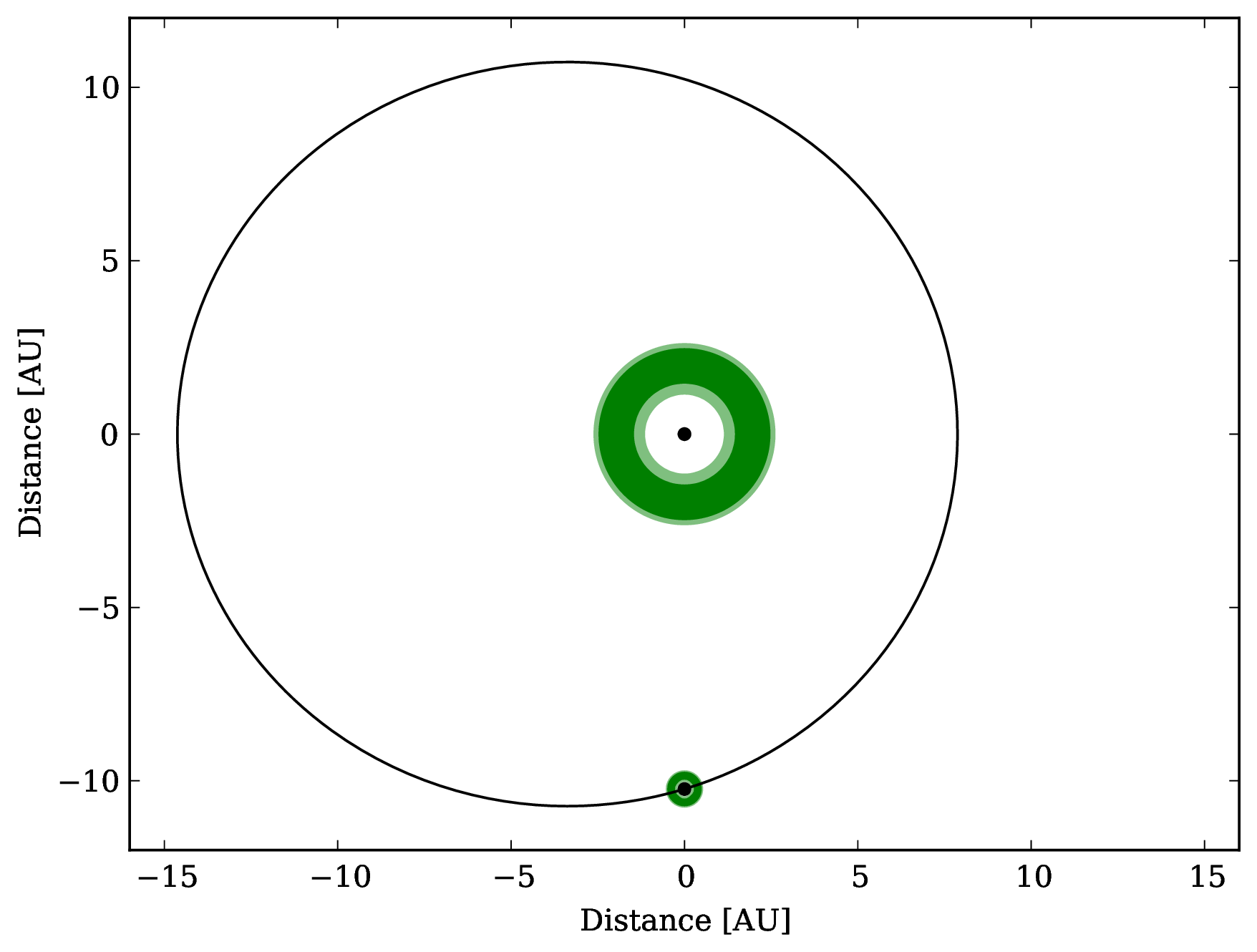}
\caption{Same as Figure 7, with the F star being the primary.}
\end{figure}

\clearpage
\begin{figure}
\vskip 1.2in
\center
\includegraphics[scale=0.8]{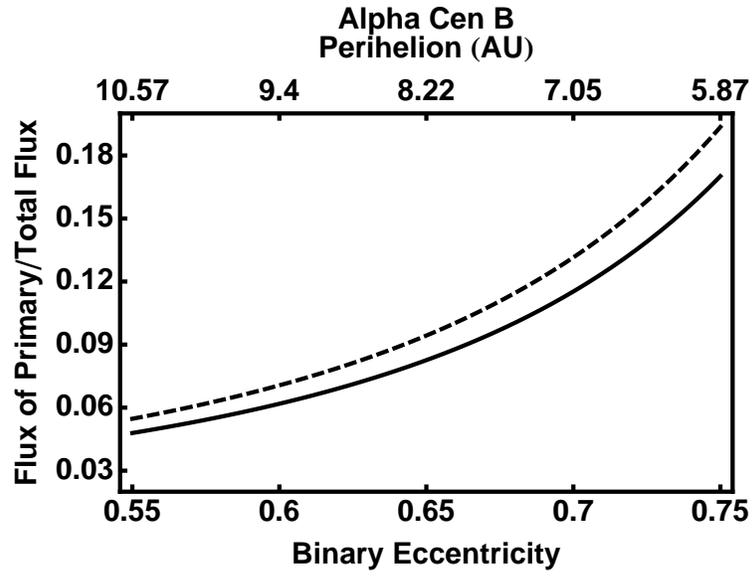}
\caption{Maximum added flux at closest distance (periastron) from a secondary from a G2V secondary at the limits of the Binary HZ of a K1V star in a generalized S-type binary similar to $\alpha$ Cen (${a_{\rm Bin}}=23.5$ AU) as a function of the binary eccentricity. The outer edge of the narrow (solid) and empirical (dashed) Binary HZs are shown.}
\end{figure}

\clearpage
\begin{figure}
\center
\vskip 1.2in
\includegraphics[scale=0.8]{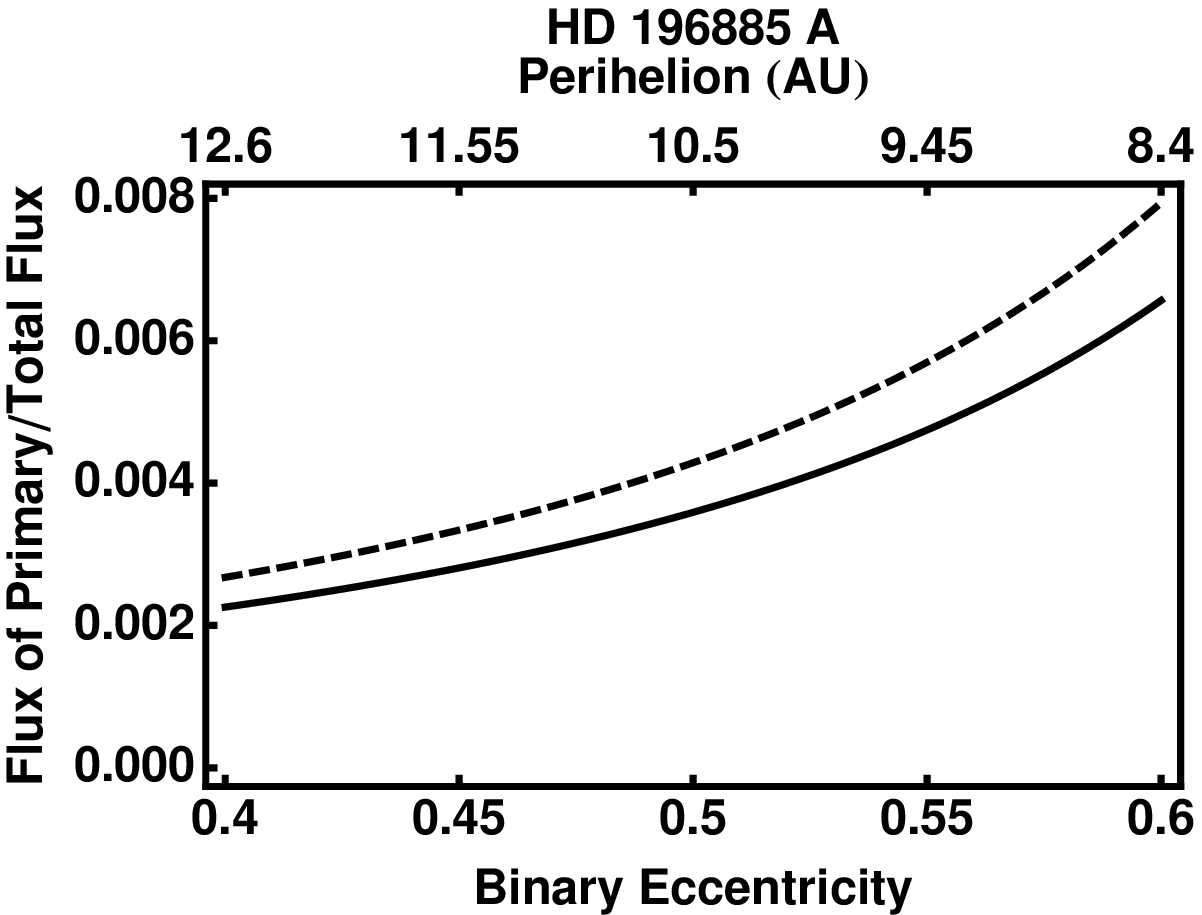}
\vskip 30pt
\includegraphics[scale=0.8]{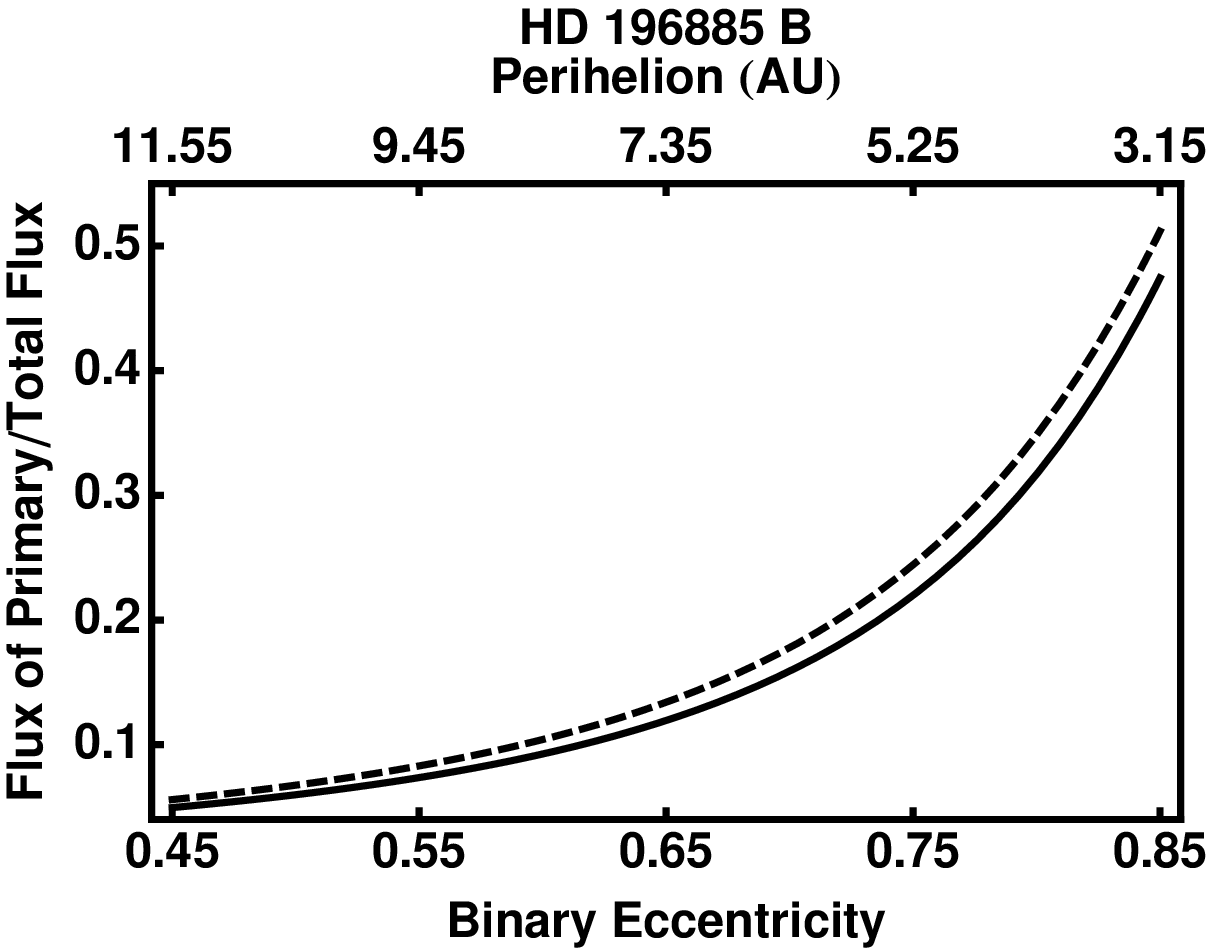}
\caption{Maximum added flux at closest distance (periastron) from a secondary at the limits of the Binary HZ of a F8V (top) and M1V (bottom) star in a generalized S-type binary similar to HD 196885, (${a_{\rm Bin}}\sim 21$ AU)  as a function of the binary eccentricity. The outer edge of the narrow (solid) and empirical (dashed) Binary HZs are shown.
}
\end{figure}

\clearpage

\begin{deluxetable}{c|c|c|c|c}
\tabletypesize{\scriptsize}
\tablecaption{Coefficients of equation (4) (Kopparapu et al. 2013b) and solar flux at the limits of the HZ.}
\label{table1}
\tablewidth{0pt}
\tablehead{\colhead{} & \multicolumn{2}{c}{Narrow HZ} & \multicolumn{2}{c}{Empirical HZ} \\
\hline
\colhead {} & \colhead{Runaway Greenhouse} & \colhead {Maximum Greenhouse} &  
\colhead{Recent Venus} & \colhead {Early Mars}
}
 \startdata
$l_{\rm {x-Sun}}$ (AU) & 0.97  & 1.67  &  0.75  &   1.77   \\
\hline
Flux (Solar Flux $@$ Earth) & 1.06  & 0.36  & 1.78 & 0.32 \\
\hline
a    & $1.2456 \times {10^{-4}}$   & $5.9578 \times {10^{-5}}$   & $1.4335 \times {10^{-4}}$   & $5.4471 \times {10^{-5}}$ \\
\hline
b    & $1.4612 \times {10^{-8}}$   & $1.6707 \times {10^{-9}}$   & $3.3954 \times {10^{-9}}$   & $1.5275 \times {10^{-9}}$ \\
\hline
c    & $-7.6345 \times {10^{-12}}$ & $-3.0058 \times {10^{-12}}$ & $-7.6364 \times {10^{-12}}$ & $-2.1709 \times {10^{-12}}$ \\
\hline
d    & $-1.7511 \times {10^{-15}}$ & $-5.1925 \times {10^{-16}}$ & $-1.1950 \times {10^{-15}}$ & $-3.8282 \times {10^{-16}}$ \\

\enddata
\end{deluxetable}


\begin{deluxetable}{l|c|c|c|c|c}
\tablecaption{Samples of Spectral Weight Factors}
\label{table1}
\tablewidth{0pt}
\tablehead{\colhead{Star} & \colhead{Eff. Temp} & \multicolumn{2}{c}{Narrow HZ} & \multicolumn{2}{c}{Empirical HZ} \\
\hline
\colhead {} & \colhead{} & \colhead{inner} & \colhead {outer} &  \colhead{inner} & \colhead {outer} 
}
\startdata
F0                   & 7300 & 0.850 & 0.815 & 0.902 & 0.806 \\
F8V (HD 196885 A)    & 6340 & 0.936 & 0.915 & 0.957 & 0.913 \\
G0                   & 5940 & 0.981 & 0.974 & 0.987 & 0.973 \\
G2V ($\alpha$ Cen A) & 5790 & 0.999 & 0.998 & 0.999 & 0.998 \\
K1V ($\alpha$ Cen B) & 5214 & 1.065 & 1.100 & 1.046 & 1.103 \\
K3                   & 4800 & 1.107 & 1.179 & 1.079 & 1.186 \\
M1V (HD 196885 B)    & 3700 & 1.177 & 1.383 & 1.154 & 1.419 \\
M5                   & 3170 & 1.192 & 1.471 & 1.179 & 1.532 \\
M2                   & 3520 & 1.183 & 1.414 & 1.163 & 1.458 \\
\enddata
\end{deluxetable}


\begin{deluxetable}{l|c|c|c|c|c|c|c|c}
\tabletypesize{\scriptsize}
\tablecaption{Estimates of the boundaries of the Binary HZ  for the max. and min. flux from the secondary
at closest and farthest approach between a fictitious planet and the secondary.}
\label{table2}
\tablewidth{0pt}
\tablehead{\colhead{Host Star} & \multicolumn{4}{c}{Estimates of Narrow HZ (AU)} & \multicolumn{4}{c}
{Estimates of Empirical HZ (AU)} \\
\hline
\colhead{} & \multicolumn{2}{c}{\underline{With Secondary}} &  \multicolumn{2}{c}{\underline{Without Secondary}} & 
\multicolumn{2}{c}{\underline{With Secondary}} &  \multicolumn{2}{c}{\underline{Without Secondary}} \\
\colhead {} & \colhead{inner} & \colhead {outer} &  \colhead{inner} & \colhead {outer}  & \colhead{inner} & 
\colhead {outer} &  
\colhead{inner} & \colhead {outer}}
 \startdata
$\alpha$ Cen A (Max) & 1.197 & 2.068 & 1.195 & 2.056 & 0.925 & 2.194 & 0.924 & 2.179\\
$\alpha$ Cen A (Min)  & 1.195 & 2.057 & 1.195 & 2.056 & 0.924 & 2.181 & 0.924 & 2.179    \\
\hline
$\alpha$ Cen B (Max) & 0.712 & 1.259 & 0.708 & 1.238 & 0.544 & 1.340 & 0.542 & 1.315\\
$\alpha$ Cen B (Min) & 0.708 & 1.241 & 0.708 & 1.238 & 0.543 & 1.317 & 0.542 & 1.315    \\
\hline
HD 196885 A (Max)  & 1.454 & 2.477 & 1.454 & 2.475 & 1.137 & 2.622 & 1.137 & 2.620 \\
HD 196885 A (Min)   & 1.454 & 2.475 & 1.454 & 2.475 & 1.137 & 2.620 & 1.137 & 2.620    \\
\hline
HD 196885 B (Max)  & 0.260 & 0.491 & 0.258 & 0.481 & 0.198 & 0.529 & 0.197 & 0.516  \\
HD 196885 B (Min)   & 0.258 & 0.483 & 0.258 & 0.481 & 0.197 & 0.518 & 0.197 & 0.516    \\
\enddata
\end{deluxetable}

\end{document}